\documentclass[12pt,a4paper]{article}
\pdfoutput=1
\usepackage{jheppub}
\usepackage{cite}

\usepackage{caption}
\usepackage{subcaption}
\usepackage{array,arydshln}
\usepackage{braket}
\usepackage{amsmath,amssymb}
\usepackage{here}
\usepackage{ascmac}

	\newcommand{\be}{\begin{equation}}
		\newcommand{\ee}{\end{equation}}
	\newcommand{\bea}{\begin{eqnarray}}
		\newcommand{\eea}{\end{eqnarray}}
	\newcommand{\beas}{\begin{eqnarray*}}
		\newcommand{\eeas}{\end{eqnarray*}}
	\newcommand{\nn}{\nonumber}

\newcommand{\pho}{F}

\title{Out-of-time-ordered correlators 
in the IP matrix model}

\author[a]{Norihiro Iizuka} 
\author[b]{and Mitsuhiro Nishida}

\affiliation[a]{Department of Physics, Osaka University, Toyonaka, Osaka 560-0043, JAPAN}
\affiliation[b]{Department of Physics, Pohang University of Science and Technology, Pohang 37673, Korea}

\emailAdd{iizuka@phys.sci.osaka-u.ac.jp} 
\emailAdd{nishida124@postech.ac.kr}

\abstract{We study the out-of-time-ordered correlators (OTOCs) in the IP matrix model \cite{Iizuka:2008hg}. It was shown in \cite{Michel:2016kwn} that OTOCs do not grow when the adjoint is massless. We generalize the analysis of OTOCs to general nonzero masses $m > 0$ for the adjoint, where we give a new prescription for analytic continuation in time such that we can evaluate OTOCs numerically using the retarded Green function. Despite the fact that the behaviors of the two-point functions, spectral density, and the Krylov complexity change drastically depending on whether the adjoint is massless or not, in the parameter ranges we study, we do not see the exponential growth of OTOCs for the massive adjoint cases. We end with a discussion of the comparison of this model with the SYK model and possible modification of the model. 
}

\begin{document}

\begin{flushright}
{\small OU-HET-1208}
 \\
\end{flushright}

\maketitle

\section{Introduction}

The IP matrix model \cite{Iizuka:2008hg} is a simple large $N$ matrix model, introduced previously as a toy model of the gauge theory dual of an AdS black hole.
It shows the key features of black holes, thermalization, and information loss as investigated in the original work. In addition, recently Krylov complexity, a new diagnostic of chaos, of the model has been proposed \cite{Parker:2018yvk} and it was shown in \cite{Iizuka:2023pov, Iizuka:2023fba} that it exhibits exponential growth. 
Since this is a nice example of exponential growth in the Krylov complexity, it is interesting to study more nature of the model. 
Note that the IP matrix model is a large $N$ matrix model. Since large $N$ matrix models are very commonplace in D-brane holography settings \cite{Banks:1996vh, Maldacena:1997re, Itzhaki:1998dd}, 
by examining various properties of large $N$ matrix models, we would like to understand better about 
what toy models allow to resum diagrams, solve, and yet produce nontrivial physics. 
The goal of this paper is to study another diagnostic of chaos, namely, out-of-time-ordered correlators (OTOCs) \cite{larkin1969quasiclassical, Lieb:1972wy, Shenker:2013pqa, Shenker:2013yza, Roberts:2014isa, Shenker:2014cwa, Kitaevtalk, Kitaev-2014, Maldacena:2015waa}, in the IP matrix model.

The IP model consists of a harmonic oscillator of the adjoint $X_{ij}$ with mass $m$ and a probe fundamental $\phi_i$ with mass $M$, and they interact through a trilinear coupling. We always consider the limit where the mass of the fundamental $M$ is much heavier\footnote{The reason why we take a large $M/T$, $M/m$ is that only then can we solve it analytically in the large $N$ limit. $M$ is a parameter corresponding to the probe distance, so when $M$ is small, it means that the probe is close enough to the black hole. In fact, when the probe is close to the black hole, it is caught in the $N^2$ dynamics of the black hole and the gauge symmetry enhancement $U(N) \times U(1) \to U(N+1)$ occurs as \cite{Iizuka:2001cw}. There it is no longer a probe.} than the mass of the adjoint, {\it i.e.,} $M \gg m$ limit and temperature,  $M \gg T = \beta^{-1}$, with various parameter ranges of $ \beta m$. 
At finite $N$, the IP matrix model consists of a finite number of interacting harmonic oscillators; thus, manifestly a unitary quantum system. However, in the large $N$ limit, this model shows thermalization and information loss. For the massless adjoint $m=0$ case, the model is rather boring and the spectrum for the fundamental shows the Wigner semi-circle law. However, for massive adjoint case $m > 0$, the spectrum for the fundamental shows a rich structure as follows. At zero temperature, the spectrum is discrete, a collection of delta functions. On the other hand at a high-temperature limit, the spectrum can become continuous and gapless and the Green function decays exponentially in time, indicating information loss \cite{Maldacena:2001kr}. How the spectrum changes from the zero-temperature discrete one to the infinite temperature continuum and gapless one is quite nontrivial.

\begin{figure}[tbp]
  \begin{minipage}[b]{0.45\linewidth}
\, \hspace{10mm}
\includegraphics[keepaspectratio, scale=0.44]{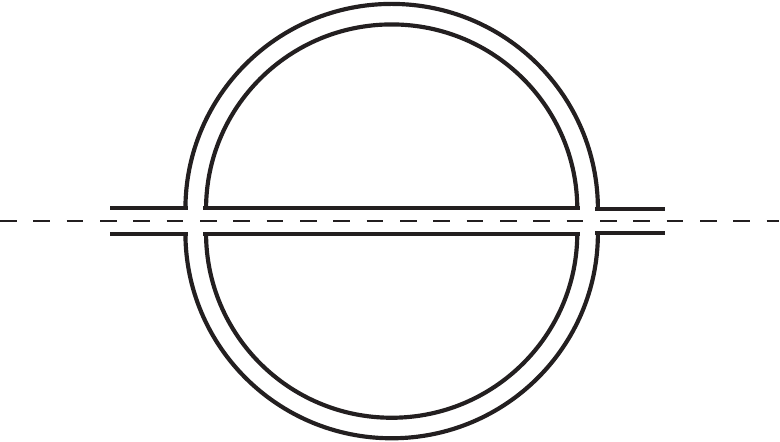}
  \end{minipage}
  \begin{minipage}[b]{0.45\linewidth}
   \,\,  \includegraphics[keepaspectratio, scale=0.7]{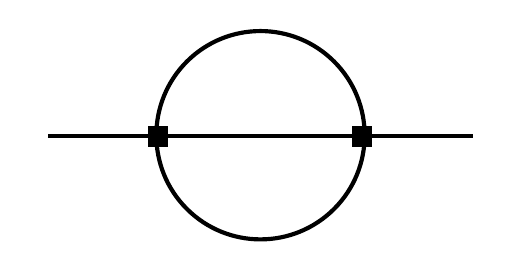}
  \end{minipage}
  \caption{Left: The basic graphical unit studied in \cite{Festuccia:2006sa}. The IP model iterates a basic unit which is just one side of this, above the dashed line. \\Right: The melonic diagram representing the corrections in the two-point function for $q=4$ SYK model \cite{Kitaev-2014}.} 
  \label{Fig:2point}
\end{figure}

\begin{figure}[tbp]
\centering
 \includegraphics[keepaspectratio, scale=0.5]{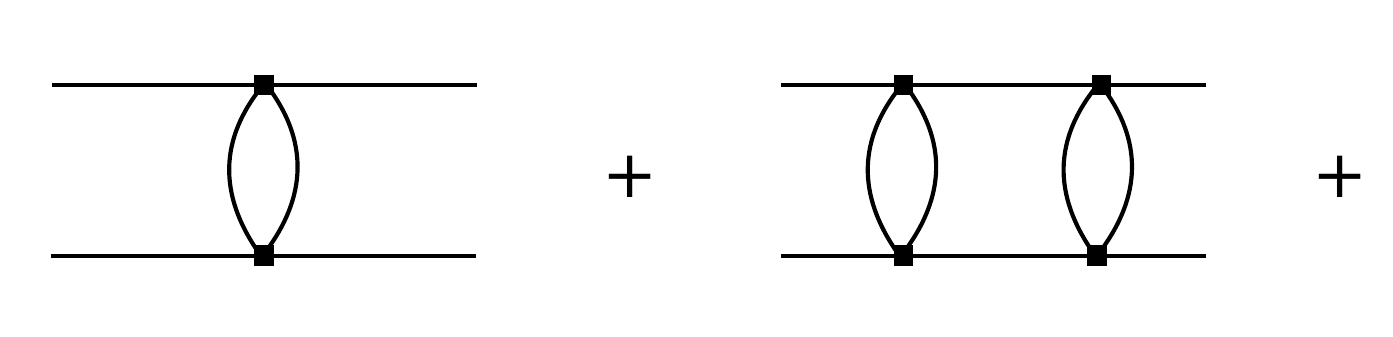}
  \caption{The ladder diagrams with one and two rungs in the SYK $q=4$ model.}
  \label{Fig:SYK4point}
 \centering
 \includegraphics[keepaspectratio, scale=0.5]{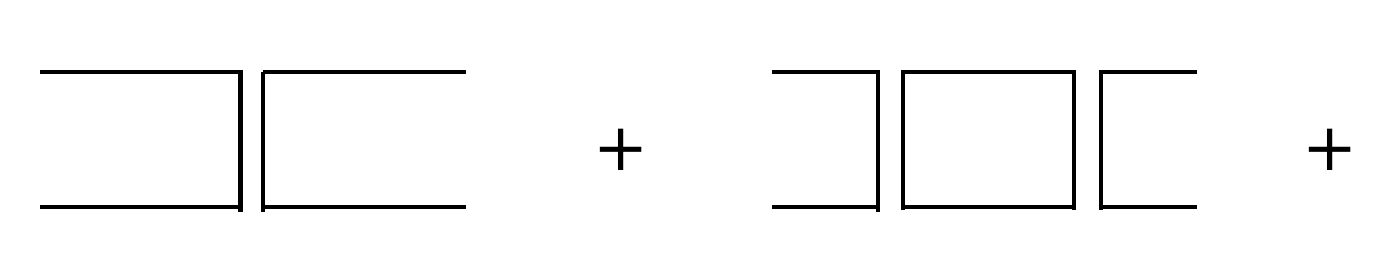}
  \caption{The planar ladder diagrams with one and two rungs in the IP model.}
      \label{Fig:IP4point0}
\end{figure}

It would be worthwhile to comment on the similarity of the Feynman diagrams between the Sachdev-Ye-Kitaev (SYK) model \cite{Polchinski:2016xgd, Kitaev-2014} and the IP matrix model. 
For the two-point function, the diagrams representing interactions are shown in Fig.~\ref{Fig:2point}.  Fig.~\ref{Fig:2point} left is the ``sunset'' diagram for the decay of the two-point function and the late time breakdown of perturbation theory, and Fig.~\ref{Fig:2point} right is the ``melonic'' diagram for the SYK $q=4$ model. 
For the out-of-time-ordered four-point function, we need to sum over ladder diagrams and Fig.~\ref{Fig:SYK4point} is the ladder diagram for the SYK model and Fig.~\ref{Fig:IP4point0} is the ladder one for the IP model. 
As is clear from Fig.~\ref{Fig:2point}, Fig.~\ref{Fig:SYK4point}, Fig.~\ref{Fig:IP4point0},  the relevant Feynman diagrams for two-point and four-point have very similar structures.
Given the similarity of the relevant diagrams, it is quite natural to ask how the IP model's OTOCs behave.

Note that there are manifest differences between the SYK model and the IP model. In the SYK there is low energy emergent conformal symmetry and its breaking, which is crucial for OTOCs to grow \cite{Kitaev-2014, Maldacena:2016hyu, Maldacena:2016upp, Polchinski:2016xgd}. On the other hand, in the IP matrix model, there is no such emergent conformal symmetry. 
Although both the SYK model and the IP model contain $N$ for the large $N$, the lack of conformal invariance in the IP model is due to the fact that there are several dimensionful parameters in the model; $\lambda$, the t' Hooft coupling, $m$, the mass of the adjoint $X_{ij}$ and $T$, the temperature for the adjoint. On the other hand, in the SYK model, we have a unique scale $J$, which is the scale set by the Gaussian randomness and at low energies, there is an emergent conformal symmetry. 

The OTOCs for the $m=0$ limit were analyzed before in \cite{Michel:2016kwn} and it was shown that for the $m=0$ case, OTOCs in the IP model do not show exponential growth. However as is seen in \cite{Iizuka:2008hg, Iizuka:2023pov, Iizuka:2023fba}, the spectral density, the late time behaviors of the two-point functions, and the Krylov complexity, all behave very differently whether we take the $m=0$ limit or not. Below is a quick summary of the difference between $m=0$ and $m > 0$ for the IP matrix model. 
\begin{itemize}
\item Regarding the spectral density at nonzero temperature $T > 0$ \cite{Iizuka:2008hg}: 
\begin{itemize}
\item for $m=0$, it is given by the Wigner's semi-circle. 
\item for $m > 0$, it is given by infinite collections of cuts with gapped spectral or continuous spectral without gap. 
\end{itemize}
\item Regarding the two-point function at the late time \cite{Iizuka:2008hg}: 
\begin{itemize}
\item for $m=0$, it is given by the power law decay in time. 
\item for $m > 0$, at high temperature, it is given by the exponential decay in time. 
\end{itemize}
\item Regarding the Krylov complexity at nonzero temperature $T > 0$ \cite{Iizuka:2023pov}:  
\begin{itemize}
\item for $m=0$, it is given by the linear growth in time. 
\item for $m > 0$, it grows exponentially in time. 
\end{itemize}
\end{itemize}
Given these manifest differences in the IP model between $m=0$ case and $m > 0$ cases, it is quite natural to see how the OTOCs behave as a function of time for $m > 0$ cases as well.   
Note that in the IOP model \cite{Iizuka:2008eb}, another matrix-vector model, OTOCs do not show exponential growth \cite{Michel:2016kwn}, and Krylov complexity does not show exponential growth as well \cite{Iizuka:2023fba}.

The organization of this paper is as follows. In section \ref{IPreview}, we briefly summarize the model, two-point function, and its spectral density. In section \ref{OTOCsection} after we review the analysis for the four-point function studied in \cite{Michel:2016kwn}, we generalize the analysis to the general $m > 0 $ cases and 
we give a new prescription for analytic continuation in time such that we can evaluate the IP model OTOCs numerically using the retarded Green function. Section \ref{mainnumerics} is for numerical analysis of the IP model OTOCs.  
We end with conclusions and discussions on possible generalizations of the model in section \ref{conclusion}.


\section{The IP matrix model two-point function and the spectral density}\label{IPreview}

\subsection{The IP matrix model}\label{sec:IP}
We briefly review the IP matrix model. For more details, see \cite{Iizuka:2008hg}. 
The IP model contains 
a Hermitian matrix field $X_{ij}(t)$ and a complex vector field $\phi_i(t)$. $X_{ij}(t)$ and $\phi_i(t)$ are harmonic oscillators with masses $m$ and $M$, in the $U(N)$ adjoint and fundamental representations, respectively. 
They obey the conventional quantization condition, 
\begin{align}
[ X_{ij}, \Pi_{kl} ] = i \delta_{il} \delta_{jk} \,, \quad  [\phi_i, \pi_j] = i \delta_{ij} \,, 
\end{align} 
and the Hamiltonian is 
\begin{eqnarray}
H = \frac{1}{2} {\rm Tr}(\Pi^2) + \frac{m^2}{2} {\rm Tr}(X^2) + M(a^\dagger a + \bar a^\dagger \bar a)  
+ g (a^\dagger X a + \bar a^\dagger X^T \bar a) + \Delta H_{\rm stab} \,, \quad \label{ham}
\end{eqnarray}
where $a^\dagger_i$ and $a_i$ are creation/annihilation operator for a vector field $\phi_i$, 
\begin{equation}
a_i = \frac{ \pi_i^\dagger  - i M\phi_i}{\sqrt{2M}}\ ,\quad
\bar a_i = \frac{ \pi_i  - i M\phi_i^\dagger}{\sqrt{2M}}\,, 
\end{equation}
and $\Delta H_{\rm stab}$ is simply a stabilizing term%
\footnote{Since the Hamiltonian commutes with the number operator $N_{\phi} = \sum_i  a_i^\dagger a_i$, $N_{\bar{\phi}} = \sum_i  \bar{a}_i^\dagger \bar{a}_i $, 
$\Delta H_{\rm stab}$ can take the form as 
\begin{equation}
\Delta H_{\rm stab} = c  \left( N_{\phi} +N_{\bar{\phi}} \right)  ( N_{\phi} + N_{\bar{\phi}} - 1)  ( N_{\phi} + N_{\bar{\phi}} - 2) \,, 
\end{equation}
which stabilizes the ground state for sufficiently large $M$ and $c$. Since $\Delta H_{\rm stab}$ vanishes for $N_\phi + N_{\bar{\phi}} = 0,1, 2$ sector which we consider, we can neglect $\Delta H_{\rm stab}$ for the rest of the paper.}.

\begin{figure}[tbp]
    \centering
\includegraphics[keepaspectratio, scale=0.8]{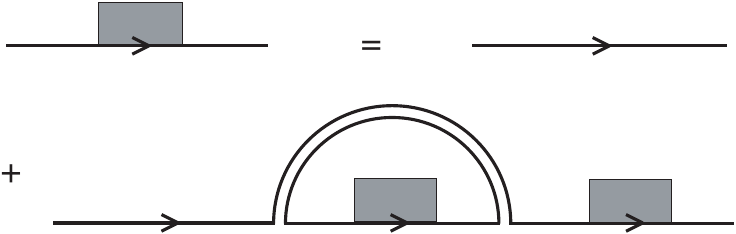}
  \caption{The planar ladder diagrams with one and two rungs in the IP model. The arrow represents the direction of the fundamental from creation operators toward annihilation operators } 
      \label{Fig:SD}
\end{figure}

We consider the following time-ordered 
Green's function for the fundamental as observable, 
\be\label{tpa}
e^{i M (t-t')} \Big{\langle} \mbox{T} \, a_i(t)\, a_j^\dagger(t') \Big{\rangle}_T := \delta_{ij} G(T, t-t') \,, 
\ee
where $T$ is temperature.  
We always consider the limit $M \gg m > 0$ and $M \gg T$. In this limit, $G(T, t-t')$ is proportional to the theta function $\theta(t-t')$ since $\beta M \to \infty$.
With 't Hooft coupling $\lambda$, the Schwinger-Dyson (SD) equation for the fundamental in the large $N$ limit becomes 
\be
\label{IPSDeq}
{G}(\omega) = {G}_0(\omega) - \lambda {G}_0(\omega) {G}(\omega) \int_{-\infty}^{\infty}  \frac{d \omega'}{2 \pi} {G}(\omega') {K}(\omega- \omega') \,, \quad \lambda := g^2 N,
\ee
where ${G}$ is a dressed propagator and ${G}_0$ is a temperature-independent bare propagator, 
\begin{align}\label{freepropagator}
{G}_0  = \frac{i}{\omega + i \epsilon} \,.
\end{align} 
$K(\omega)$ is a thermal propagator for $X_{ij}$, given by
\begin{align} 
K(\omega) = \frac{i}{1 - y} \left( \frac{1}{\omega^2 - m^2 + i\epsilon}
- \frac{y}{\omega^2 - m^2 - i\epsilon} \right)  \,,  \quad y:= e^{- m/T} \,, \label{ktherm} 
\end{align}
which is free since the backreaction of $X_{ij}$ on $\phi_i$ is suppressed by $1/N$. See Figure \ref{Fig:SD}. This model is similar to 't Hooft's two-dimensional QCD model \cite{tHooft:1974pnl} in the sense that the adjoints have no self-interaction. 

Using the fact that 
the time-ordered correlator $G(\omega)$ has no pole in upper half $\omega$-plane in the limit $\beta M \to \infty$,  
by closing the contour in the upper half-plane, we have 
\begin{align}
{G}(\omega) = {G}_0(\omega) - \frac{\lambda}{2 m (1- y)} {G}_0(\omega) {G}(\omega) \Bigl( G(\omega - m ) + y \, G(\omega + m) \Bigr)  \,.
\label{SDeq29}
\end{align}
The term $G(\omega - m )$ comes from the process where the adjoint with its energy $m$ is created. On the other hand, the term $G(\omega + m )$ is weighted by thermal factor $y$ because this term comes from the process where the adjoint with its energy $m$ is created in a thermal bath and absorbed into the fundamental. 
This equation can be written as 
\begin{align}
\hspace{-9mm} G(\omega - m)  - \frac{4}{\nu_T^2}\frac{1}{  G(\omega)} + e^{- m/T}  G(\omega + m) 
= \frac{4 i \omega}{\nu_T^2}  \label{sde4} \,,
\end{align}
where 
\begin{align}
\label{defnuT}
\nu_T^2 := \frac{\nu^2}{1- e^{- m/T}} = \frac{\nu^2}{1-y} \,, \quad \nu^2 := \frac{2\lambda}{m} \,, 
\end{align}
In this paper, we focus on the limit where $\nu_T$ = fixed. Then taking the variation of eq.~\eqref{sde4} for $y$, we have\footnote{On the other hand, if we take the other limit where $\nu^2 = \lambda/m$ = fixed, then we have 
\begin{align}
\hspace{-9mm} \partial_y G(\omega - m)  + \frac{4}{\nu_T^2}\frac{\partial_y {G(\omega)} }{(G(\omega))^2} + \partial_y G(\omega + m) + y  \, \partial_y G(\omega + m) 
= -\frac{4 }{\nu^2} \left( i \omega + \frac{1}{G(\omega)} \right)  \,.
\end{align} 
} 
\begin{align}
\hspace{-9mm} \partial_y G(\omega - m)  + \frac{4}{\nu_T^2}\frac{\partial_y {G(\omega)} }{(G(\omega))^2} + \partial_y G(\omega + m) + y  \, \partial_y G(\omega + m) 
= 0 \,. 
\label{numericalSD}\end{align} 
Starting with $y=0$ ($\leftrightarrow T=0$) analytic solution, one can obtain finite temperature Green function by solving eq.~\eqref{numericalSD}. 
The IP model spectral density drastically changes whether we take $m=0$ or $m > 0$.  

\subsection{Massless adjoint case $m=0$}
Let us first consider the case $m=0$. Then SD equation \eqref{sde4} can be solved as 
\begin{equation}
G(\omega ) = \frac{2 i}{\omega + \sqrt{\omega^2 - 2 \nu_T^2}} \,.
\label{mzeroGomega} 
\end{equation}
Since the real part of ${G}$ is a spectral density, defining 
\begin{align}
{\rm Re}\, G(T,\omega) =  \pi \pho(\omega), 
\end{align}
$\pho(\omega)$ is the spectral density. Thus, for $m=0$ case, the spectral density is simply given by a single Wigner's semi-circle law as 
\begin{align}
 \pho(\omega) = \frac{1}{\pi \nu_T^2}{\rm Re}\, \left[ \sqrt{2 \nu_T^2 - \omega^2} \right]\,.
\end{align}
Thus for the $m=0$ case, the two-point function $G(T, t)$, which is given by the Fourier transform of eq. \eqref{mzeroGomega}, decays only by the power law and we will not obtain any interesting large $N$ transition by changing the temperature. This is because we have only one scale $\nu_T$ in this massless limit. 

\subsection{Double-scaling limit $m\to 0$, $T \to 0$, $y$ = fixed}
Let's generalize the previous $m =0$ case a bit and consider the situation where we take a double scale limit where $m \to 0$, $T \to 0$ with the ratio $m/T$ of the two is fixed.
In this case, the SD equation \eqref{sde4} gives 
\begin{equation}
G(\omega ) = \frac{2 i}{\omega + \sqrt{\omega^2 - (1+y) \nu_T^2}}  \,. 
\label{doublezeroGomega} 
\end{equation}
Again it is Wigner's semi-circle law with its radius $\sqrt{1 + y} \, \nu_T$. 

\subsection{Massive adjoint case $m > 0$, $T > 0$}
On the other hand, for the $m > 0$ case, the spectral density shows more richer structure. 
Since eq.~\eqref{sde4} reduces to 
\begin{equation}
\pho(\omega - m)  - \frac{4}{\nu_T^2 |  G(T,\omega) |^2} {\pho(\omega)} + e^{- m/T}
\pho(\omega + m) = 0 \,,
\label{real} 
\end{equation}
we can see the following structure of the spectral density at nonzero temperature $T > 0$. Due to the positivity of the spectral density $\pho(\omega)$, 
if $\pho(\omega_0) =0$ at some $\omega = \omega_0$, then $\pho(\omega_0 \pm m) =0$ as well. 
Thus, if there is a gap, then it continues unboundedly by shifting $\pm m$ for $\omega$.  
We can see that if there is a cut, then that cut is accompanied by an unbounded series of additional cuts in both positive and negative $\omega$ directions. Suppose the branch cuts continue only up to $\omega = \omega_0$. In other words, suppose that 
\begin{align}
\pho(\omega_0) \neq 0  \,, \quad  \mbox{but} \quad \pho(\omega_0 + m) = 0  \, \quad \mbox{at some $\omega_0$}
\end{align}  
Then setting $\omega = \omega_0 + m $ in \eqref{real},  we obtain 
\begin{align}
\pho(\omega_0 ) + e^{- m/T}
\pho(\omega_0 + 2 m) = 0 \,.
\end{align}
However, due to the positivity of $\pho(\omega_0 )$, this immediately implies $\pho(\omega_0 )  = 0$, which is in contradiction. Similarly one can show that there is no bound on the lower $\omega$ direction as well. Thus, each cut is always accompanied by an unbounded series of additional cuts. 
Note that the structure that there is an infinite amount of $m$-translated cut exists only at nonzero temperature.  $T=0$ is special since there are poles and $|  G(T,\omega) |$ diverges. But at nonzero temperature $T > 0$, poles disappear.

\begin{figure}[thb]
\centering
     \begin{subfigure}[b]{0.44\textwidth}
         \centering
         \includegraphics[width=\textwidth]{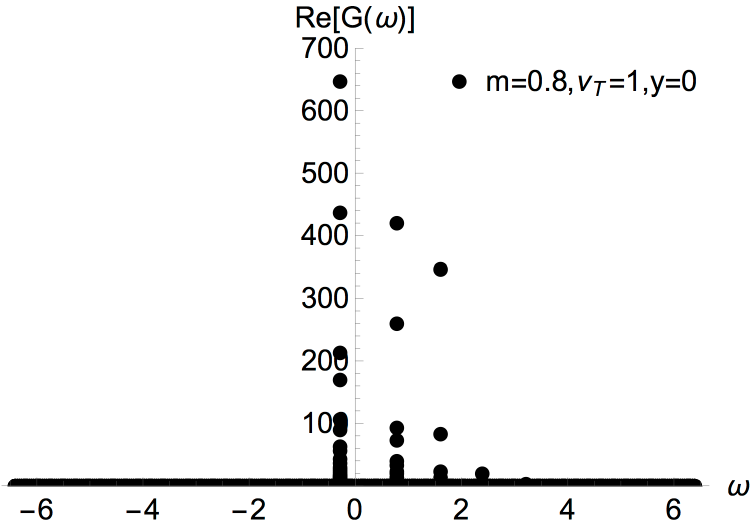}
     \end{subfigure}
\centering
     \begin{subfigure}[b]{0.44\textwidth}
         \centering
         \includegraphics[width=\textwidth]{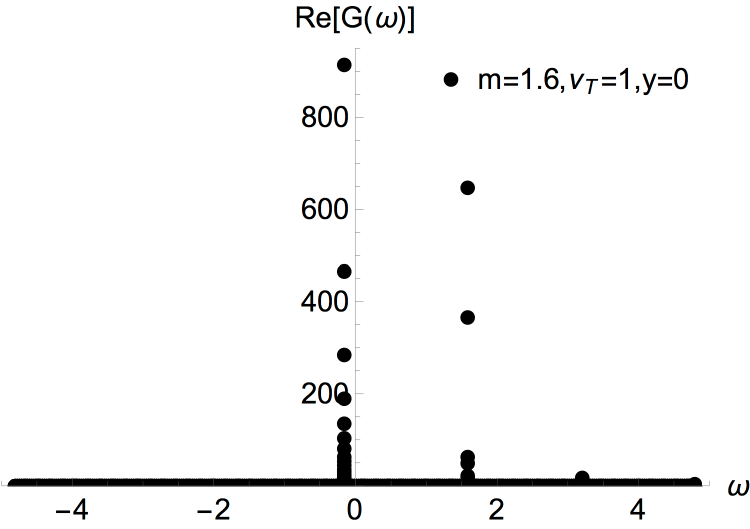}
     \end{subfigure}
     \centering
          \begin{subfigure}[b]{0.44\textwidth}
         \centering
         \includegraphics[width=\textwidth]{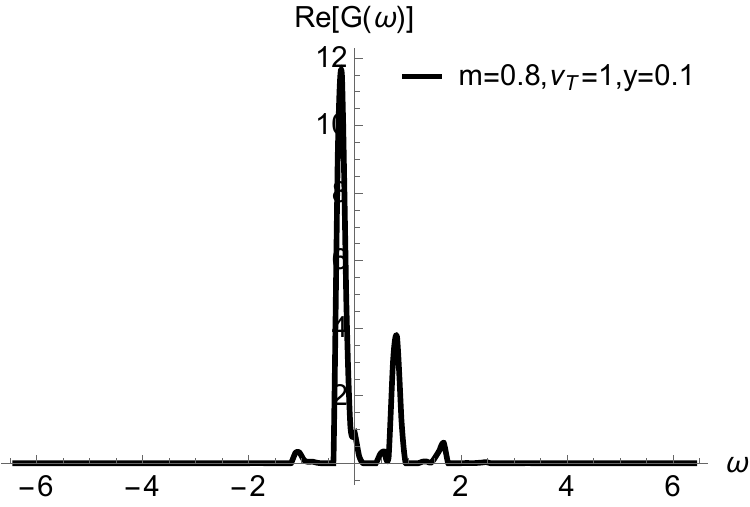}
     \end{subfigure}
     \begin{subfigure}[b]{0.44\textwidth}
         \centering
         \includegraphics[width=\textwidth]{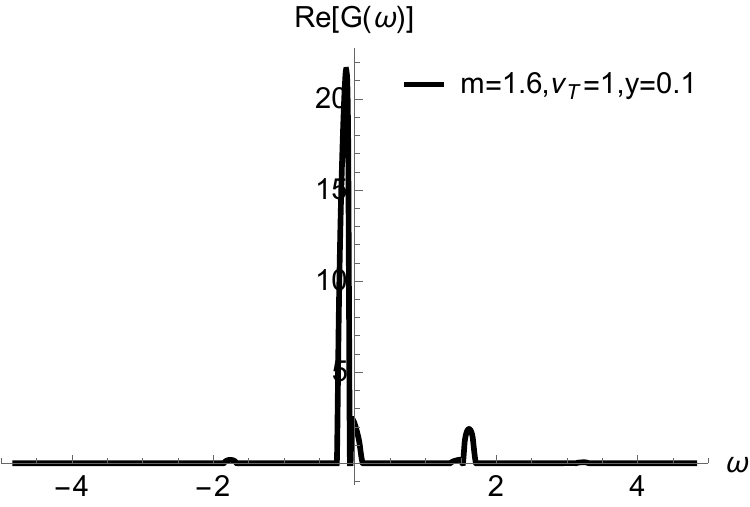}
     \end{subfigure}
        \begin{subfigure}[b]{0.44\textwidth}
         \centering
         \includegraphics[width=\textwidth]{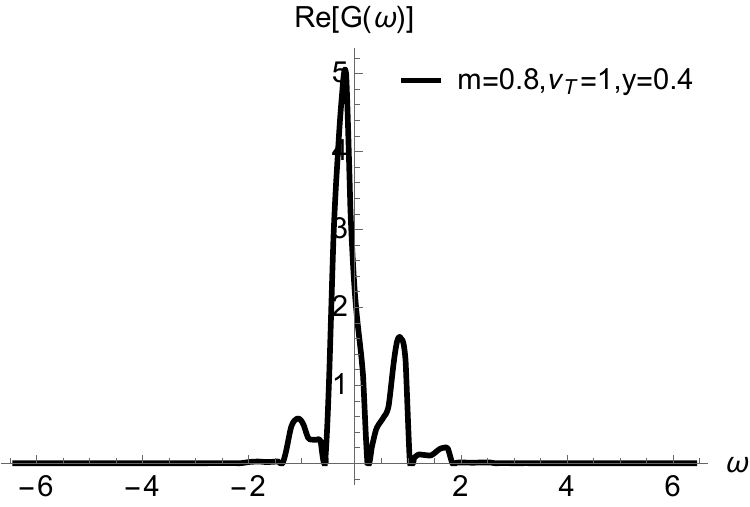}
     \end{subfigure}
         \begin{subfigure}[b]{0.44\textwidth}
         \centering
         \includegraphics[width=\textwidth]{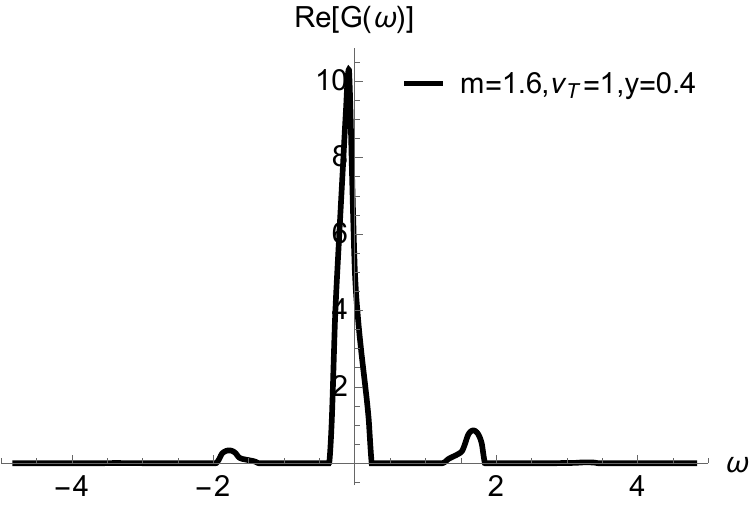}
     \end{subfigure}
       \begin{subfigure}[b]{0.44\textwidth}
         \centering
         \includegraphics[width=\textwidth]{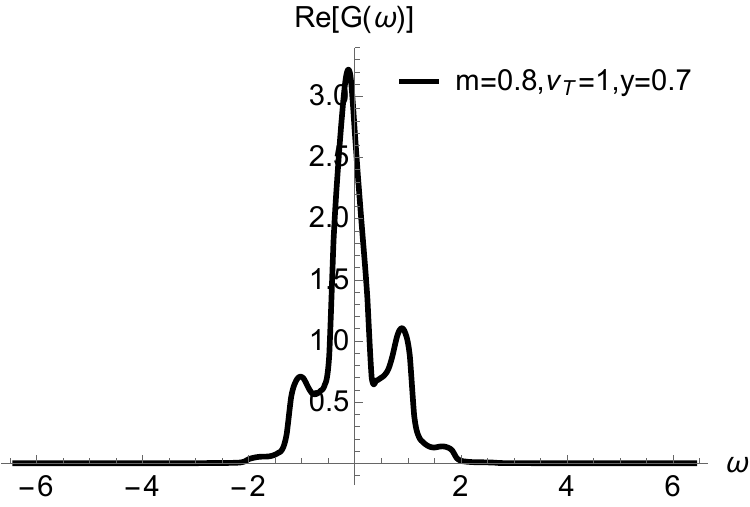}
     \end{subfigure}
            \begin{subfigure}[b]{0.44\textwidth}
         \centering
         \includegraphics[width=\textwidth]{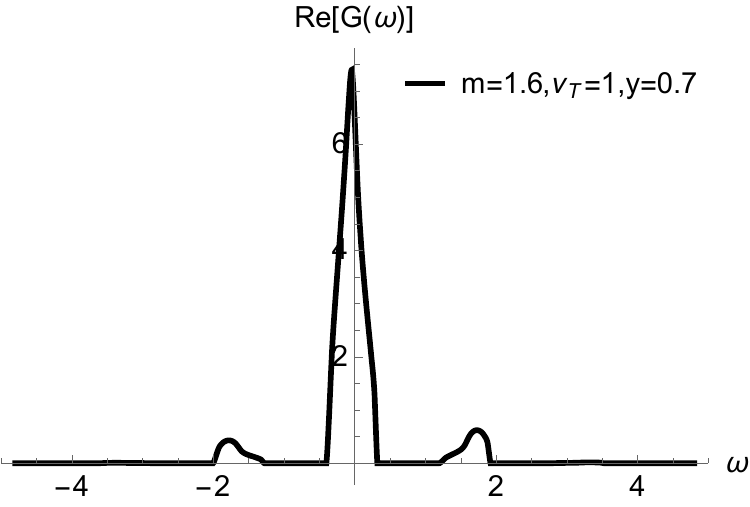}
     \end{subfigure}
       \begin{subfigure}[b]{0.44\textwidth}
         \centering
         \includegraphics[width=\textwidth]{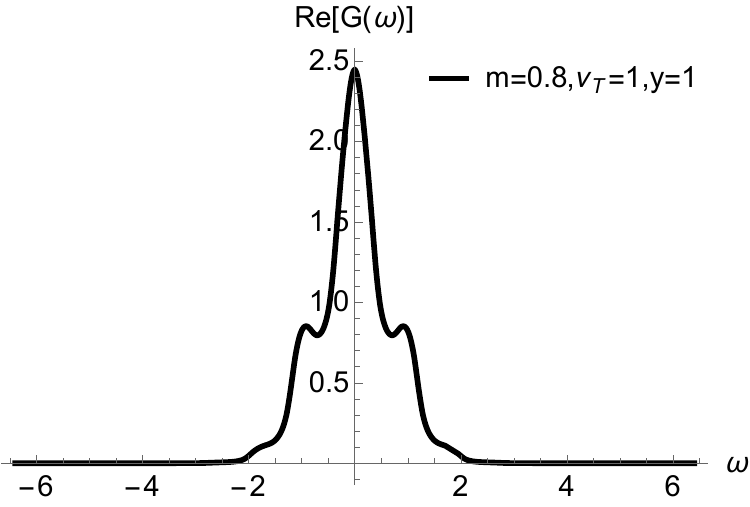}
     \end{subfigure}
       \begin{subfigure}[b]{0.44\textwidth}
         \centering
         \includegraphics[width=\textwidth]{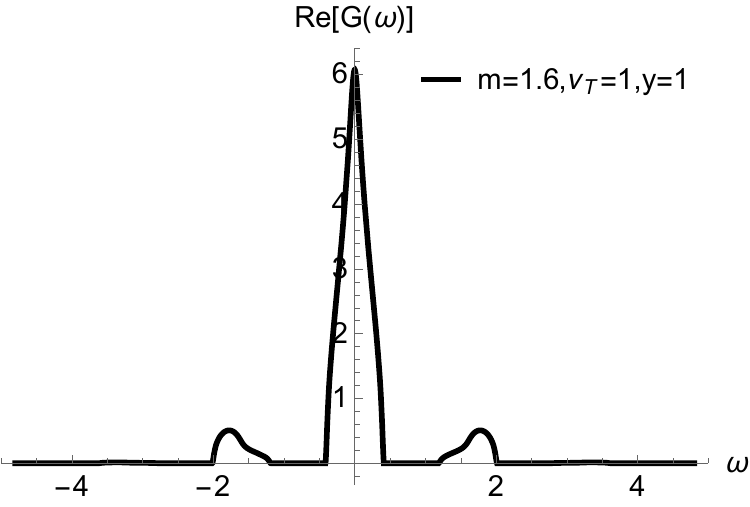}
     \end{subfigure}
                 \caption{Numerical plots of $\text{Re}[G(\omega)]$ for $\nu_T = 1$ with $m=0.8$ and $m= 1.6$ at various temperatures from $y=0$ to $y=1$, where $y=e^{-m/T}$.}
        \label{fig:ReG(w)}
\end{figure}

The IP model spectrum for $m \neq 0$ changes drastically from collections of discrete poles at $T=0$ to continuum spectrum at nonzero temperature. In Figure \ref{fig:ReG(w)}, the spectrum densities at various temperatures are shown by solving eq.~\eqref{numericalSD} numerically. For Figure \ref{fig:ReG(w)}, we set $\epsilon$ in (\ref{freepropagator}) as $\epsilon=10^{-5}$. See \cite{Iizuka:2008hg, Iizuka:2023pov} for detail. Here we comment $T=0$ and $T > 0 $ results. 

\begin{itemize}
\item

At zero temperature,   
one can solve this equation analytically by mapping the equation to the Bessel recursion relation based on the canonical calculation \cite{Iizuka:2008hg},  
\begin{align}\label{GzeroT}
 G(\omega)  
= -\frac{2i}{\nu}\frac{ J_{-\omega/m}(\nu/m)}{J_{-1-\omega/m}(\nu/m)}\ .
\end{align}
where $J$ is a Bessel function.   
The spectrum is determined by the pole of $ G(\omega) $ which is discrete for nonzero $m > 0$. There are infinite poles, which are determined by the zeros of the denominator. 
Thus any intervals between the poles at zero temperature are different from multiples of $m$ generally. 
\item As we increase the temperature, these poles become narrow cuts. However, at nonzero temperature, these cuts must obey a shift symmetry between $\omega$ and $\omega+m$ due to eq.~(\ref{real}). To satisfy this $m$-shift symmetry at nonzero temperature, additional tiny cuts appear away from the poles by $m$-shift, which can be seen from our numerical plots of $y=0.1$.
\item
In our numerical computations at infinite temperature $y=1$, the spectrum for $m=0.8$ is gapless. On the other hand, the spectrum for $m=1.6$ is gapped.  

\item As $T$ increases, the behavior of a two-point function 
\begin{align}
G(T,t)=\int_{-\infty}^{\infty}\frac{d\omega}{2\pi} G(\omega)e^{-i\omega t}
\end{align} 
changes drastically. For that, see the summary table Figure 18 in \cite{Iizuka:2023fba}. 

\end{itemize}

\newpage

\clearpage

\section{The IP matrix model four-point function}\label{OTOCsection}
\subsection{Time-ordered and out-of-time-ordered four-point functions}

\begin{figure}[htbp]
\hspace{-10mm}\,\,\,\,\,\,\, \includegraphics[keepaspectratio, scale=0.52]{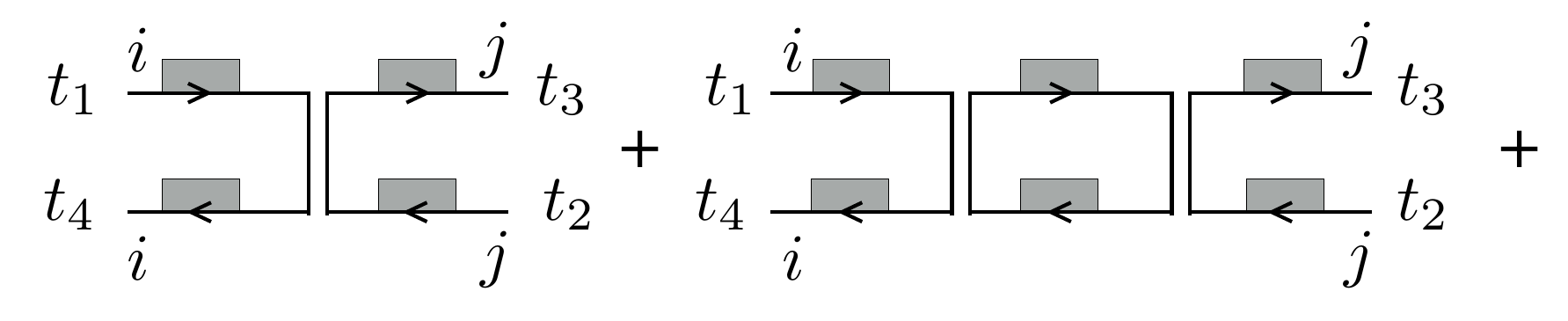}
  \caption{The planar ladder diagrams with one and two rungs in the IP model. The arrow represents the direction of the fundamental from creation operators toward annihilation operators.} 
      \label{Fig:IP4point}
\end{figure}

Let us study the connected four-point function $G_4(\omega_1, \omega_2, \omega_3, \omega_4)$. In the planar limit, it consists of a sum of ladder diagrams. In the $m=0$ limit, it is worked out in \cite{Michel:2016kwn}. Here we will work out the general nonzero $m >  0$ case. We basically take the same convention as \cite{Michel:2016kwn}, thus we set the ingoing momenta, $\omega_1, \omega_2$, while the outgoing momenta, $\omega_3, \omega_4$ as shown in Fig.~\ref{Fig:IP4point}. 
As in the case of the two-point function, to proceed we must work in the limit where $\nu_T$ is fixed.  
Thus $G_4(\omega_1, \omega_2, \omega_3, \omega_4)$ can be decomposed into subsets which consists of $n$ rungs, 
\be
{G}_4(\omega_1, \omega_2, \omega_3, \omega_4) = \sum_{n=1}^\infty {G}^{(n)}_4(\omega_1, \omega_2, \omega_3, \omega_4) \,.
\ee

Consider the ladder diagram that consists of a single rung, $G_4^{(1)}$, is given by,
\begin{align} \label{eq:single}
G_4^{(1)} = (-i g)^2 \int \frac{dp_1}{2\pi} G(\omega_1) G(\omega_1 - p_1) G(\omega_2) G(\omega_2+p_1) {K}(p_1)\nn \\
\times \left( 2 \pi \right)^2 \delta(w_1- \omega_3 - p_1 ) \delta(w_2 - \omega_4 + p_1 )  \,,
\end{align}
where the thermal propagator ${K}(p_1)$ is given by eq.~\eqref{ktherm}. 
Similarly, the ladder diagram that consists of two rungs yields
\begin{align} \label{eq:two}
G_4^{(2)} &= N(-i g)^4 \int \frac{dp_1}{2\pi}\frac{dp_2}{2\pi} G(\omega_1) G(\omega_1 - p_1) G(\omega_1 -  \sum_{i=1}^2 p_i) G(\omega_2)  G(\omega_2+p_2) \nn \\
& \quad  \times  G(\omega_2+ \sum_{i=1}^2 p_i)  K(p_1)K(p_2)  \left( 2 \pi \right)^2 \delta(w_1- \omega_3 - \sum_{i=1}^2 p_i ) \delta(w_2 - \omega_4 + \sum_{i=1}^2 p_i ).
\end{align}
In general, the ladder diagram that consists of $n$ rungs yields
\begin{align} \label{eq:n}
& G_4^{(n)} = N^{n-1}(-i g)^{2n} \int \frac{dp_1 \cdots dp_n}{(2\pi)^n}   G(\omega_1) G(\omega_1 - p_1) G(\omega_1 - p_1-p_2) \cdots G(\omega_1 - \sum_{i=1}^n p_i)  \nn \\
& \,\, \qquad \times G(\omega_2) G(\omega_2+p_n)G(\omega_2+p_n+p_{n-1}) \cdots  G(\omega_1 + \sum_{i=1}^n p_i) K(p_1)K(p_2) \cdots K(p_n)~ \nn \\
& \,\,\, \qquad \times \left( 2 \pi \right)^2 \delta(w_1- \omega_3 - \sum_{i=1}^n p_i ) \delta(w_2 - \omega_4 + \sum_{i=1}^n p_i ) \,.
\end{align}

Given these, the four-point function in position space can be obtained by its Fourier transformation as 
\be
\label{n-thGrealtime}
{G}^{(n)}_4 (t_1, t_2, t_3, t_4) = \int  \frac{d\omega_1}{2\pi}   \frac{d\omega_2}{2\pi}   \frac{d\omega_3}{2\pi}   \frac{d\omega_4}{2\pi}  {G}^{(n)}_4  (\omega_1, \omega_2, \omega_3, \omega_4) e^{i\omega_1t_1+i\omega_2t_2-i\omega_3 t_3-i \omega_4 t_4} \,.
\ee
We first consider the time-ordered four-point function, for that purpose, we will set 
\be
\label{time-ordered}
t_3 > t_1 \,, \quad t_4 > t_2 \,.
\ee
Later we consider out-of-time-ordered correlators, for that, we will set 
\be
\label{out-of-time-ordered}
t_3 > t_1 \,, \quad t_4 < t_2 \,.
\ee
However, naively setting eq.~\eqref{out-of-time-ordered} into eq.~\eqref{n-thGrealtime}, we will see 
\be
{G}^{(n)}_4 (t_1, t_2, t_3, t_4) = 0  \quad \mbox{for} \quad t_4 < t_2 \,.
\ee
This is because to evaluate ${G}^{(n)}_4 (t_1, t_2, t_3, t_4)$ in eq.~\eqref{n-thGrealtime}, we use its Fourier transformation, $G_4^{(n)}$ in \eqref{eq:n}. However $G_4^{(n)}$ in \eqref{eq:n} are consisted of the retarded Green function $G(\omega)$, and thus, ${G}^{(n)}_4 (t_1, t_2, t_3, t_4)$ is proportional to theta function $ \theta(t_4 - t_2)$,  
\be
{G}^{(n)}_4 (t_1, t_2, t_3, t_4)  \propto \theta(t_3 - t_1)   \theta(t_4 - t_2) \,.
\ee
To see this theta function more explicitly, let us substitute eq.~\eqref{eq:n} into eq.~\eqref{n-thGrealtime} and evaluate eq.~\eqref{n-thGrealtime}. Then the four-point function ${G}^{(n)}_4 (t_1, t_2, t_3, t_4)$ becomes 
\begin{align}
&{G}^{(n)}_4 (t_1, t_2, t_3, t_4)  = N^{n-1}(-i g)^{2n}   \int \frac{dp_1 \cdots dp_n}{(2\pi)^n}   {e^{i \sum_{i=1}^n  p_i (t_3 -t_4)} } K(p_1)K(p_2) \cdots K(p_n) \nn \\
&\,\, \times \int  \frac{d\omega_1}{2\pi}  G(\omega_1) G(\omega_1 - p_1) G(\omega_1 - p_1-p_2) \cdots G(\omega_1 - \sum_{i=1}^n p_i)  e^{- i \omega_1 (t_3 - t_1)} \nn \\
& \,\, \times \int  \frac{d\omega_2}{2\pi}   G(\omega_2) G(\omega_2+p_n)G(\omega_2+p_n+p_{n-1}) \cdots  G(\omega_1 + \sum_{i=1}^n p_i)  e^{- i \omega_2 (t_4 - t_2)}  \,.
\end{align}
Because of the limit we consider, $M \gg T$, the dressed propagator $G(\omega)$ has singularities only in the lower half $\omega$-plane. Thus, $\omega_1$ and $\omega_2$ integral vanishes unless $t_3 > t_1$ and $t_4 > t_2$. In other words, by going to the upper half-plane, the integral picks up no poles, and thus it vanishes. 

Thus to allow analytic continuation such that we can obtain the OTOCs, we will add the complex conjugate of the $\omega_2$ integrand as, 
\begin{align}
& \int  \frac{d\omega_2}{2\pi}   G(\omega_2) G(\omega_2+p_n)G(\omega_2+p_n+p_{n-1}) \cdots  G(\omega_1 + \sum_{i=1}^n p_i)  \, e^{- i \omega_2 (t_4 - t_2)}  \nn \\
\to &  \int  \frac{d\omega_2}{2\pi}  \left[  G(\omega_2) G(\omega_2+p_n)G(\omega_2+p_n+p_{n-1}) \cdots  G(\omega_1 + \sum_{i=1}^n p_i)  \right. \nn \\ 
&\qquad  \left. +\,  G^*(\omega_2) G^*(\omega_2+p_n)G^*(\omega_2+p_n+p_{n-1}) \cdots  G^*(\omega_1 + \sum_{i=1}^n p_i)   \right] e^{- i \omega_2 (t_4 - t_2)} \,,
\end{align}
where $^*$ implies complex conjugate. 
Note that this prescription will not change the behavior for $t_4 > t_2$. However, due to the additional complex conjugate $G^*(\omega_2) \cdots $ terms, it allows analytic continuation to $t_2 > t_4$ case.

Thus, for time-ordered cases, $t_4 > t_2$, the newly added $G^*(\omega_2) G^*(\omega_2+p_n) \cdots$ term will not contribute since $G^*(\omega_2) G^*(\omega_2+p_n) \cdots$ term has no singularities in the lower half $\omega_2$ plane. For out-of-time-ordered cases, $t_2 > t_4$, the original $G(\omega_2) G(\omega_2+p_n) \cdots$ term will not contribute since $G(\omega_2) G(\omega_2+p_n) \cdots$ term has no singularities in the upper half $\omega_2$ plane.   

Physically taking this complex conjugate for the $\omega_2$ integrand corresponds to flipping the time direction on the bottom edge of the ladder diagrams in 
Fig.~\ref{Fig:IP4point}, 
since we obtain $\theta(t_2 - t_4)$ rather than $\theta(t_4 - t_2)$. In this case, since time is running backwards on the bottom edge, the interactions on the bottom come with a factor of $+ i g$, instead of $- i g$ \cite{Michel:2016kwn} in Ferynman diagrams. Thus the parameter for the out-of-time-ordered case must change from $(- i g)^2 \to g^2$ or equivalently $- \lambda \to + \lambda$ compared with the time-ordered case, 
and thus we reach the following prescription. \\

 In summary, we propose the following prescription for OTOCs in the IP model as follows. 
\newpage

\begin{itembox}[l]{\underline{Prescription for analytic continuation in the IP matrix model}}

Writing the four-point function as 
\begin{align}
N {G}_4 (t_1, t_2, t_3, t_4) = N \sum_{n=1}^\infty {G}^{(n)}_4 (t_1, t_2, t_3, t_4)  \,,
\end{align}
\vspace{-3mm}
\begin{itemize}
\item For the time-ordered four-point function,  where time is set as eq.~\eqref{time-ordered}, 
we have 
\begin{align}
&\hspace{-5mm}N {G}^{(n)}_4  (t_1, t_2, t_3, t_4)  = (- \lambda)^{n}   \int \frac{dp_1 \cdots dp_n}{(2\pi)^n}   {e^{i \sum_{i=1}^n  p_i (t_3 -t_4)}}  K(p_1)K(p_2) \cdots K(p_n) \nn \\
&\hspace{-5mm}\quad \times \int  \frac{d\omega_1}{2\pi}  G(\omega_1) G(\omega_1 - p_1) G(\omega_1 - p_1-p_2) \cdots G(\omega_1 - \sum_{i=1}^n p_i) e^{- i \omega_1 (t_3 - t_1)} \nn \\
&\hspace{-5mm} \quad \times \int  \frac{d\omega_2}{2\pi}   G(\omega_2) G(\omega_2+p_n)G(\omega_2+p_n+p_{n-1}) \cdots  G(\omega_1 + \sum_{i=1}^n p_i)  e^{- i \omega_2 (t_4 - t_2)}  \,.
\label{timeorder}
\end{align}
\item For the out-of-time-ordered four-point function, where time is set as eq.~\eqref{out-of-time-ordered}, we have 
\begin{align}
&\hspace{-3mm} N {G}^{(n)}_4 (t_1, t_2, t_3, t_4)   =  \left( +\lambda\right)^{n}   \int \frac{dp_1 \cdots dp_n}{(2\pi)^n}   {e^{i \sum_{i=1}^n  p_i (t_3 -t_4)} } K(p_1)K(p_2) \cdots K(p_n) \nn \\
&\hspace{-8mm}\quad \times \int  \frac{d\omega_1}{2\pi}  G(\omega_1) G(\omega_1 - p_1) G(\omega_1 - p_1-p_2) \cdots G(\omega_1 - \sum_{i=1}^n p_i) e^{- i \omega_1 (t_3 - t_1)}  \nn \\
&\hspace{-8mm} \quad \times \int  \frac{d\omega_2}{2\pi}   G^*(\omega_2) G^*(\omega_2+p_n)G^*(\omega_2+p_n+p_{n-1}) \cdots  G^*(\omega_1 + \sum_{i=1}^n p_i)  e^{- i \omega_2 (t_4 - t_2)} \,.
\label{prescription}
\end{align}
\end{itemize}
\end{itembox}

For the time-ordered one, we must handle the $p$ integral which is not easy. However, for the out-of-time-ordered one, we can do the $p_i$ integration fortunately in the lower half $p_i$ plane since neither $G(\omega - p_i )$ nor $G^*(\omega + p_i)$ has singularities in the lower half $p_i$ plane. 

Before we proceed with the out-of-time-ordered case, let us work out explicit examples to check the consistency of our prescription. For that purpose we first review the $m=0$ case studied in \cite{Michel:2016kwn}.

\subsection{Consistency check for $m=0$ case}

For the case where the adjoint is massless, {\it i.e.,} $m=0$, analytical calculations are possible as shown in \cite{Michel:2016kwn}. To check the consistency of the prescription eq.~\eqref{prescription}, let us evaluate the time-ordered four-point function from eq.~\eqref{timeorder} and then make analytic continuation of time and compare with eq.~\eqref{prescription}. 

In the massless limit $m\to 0$, the adjoint propagator reduces to 
\begin{align}
K(p) \to \frac{2 \pi}{m (1-y)} \delta(p)
\end{align}
where $y= e^{- m/T}$. 
The $n$ rung contribution $N {G}^{(n)}_4 (t_1, t_2, t_3, t_4)$ for the time-ordered four-point function eq.~\eqref{timeorder} becomes 
\begin{align}
N {G}^{(n)}_4 (t_1, t_2, t_3, t_4)  &= \left( - \frac{\lambda}{m (1-y)} \right)^{n}   
\int  \frac{d\omega_1}{2\pi}  \left( G(\omega_1) \right)^{n+1}   e^{- i \omega_1 (t_3 - t_1)} \nn \\ 
&  \qquad \times \int  \frac{d\omega_2}{2\pi}   \left( G(\omega_2) \right)^{n+1}  e^{- i \omega_2 (t_4 - t_2)}   \,.
\label{zeromasstimeorder}
\end{align}

On the other hand, the $n$ rung contribution $N {G}^{(n)}_4 (t_1, t_2, t_3, t_4)$ for the out-of-time-ordered four-point function eq.~\eqref{prescription} becomes 
\begin{align}
N {G}^{(n)}_4 (t_1, t_2, t_3, t_4) & = \left( \frac{\lambda}{m (1-y)} \right)^{n}   
\int  \frac{d\omega_1}{2\pi}  \left( G(\omega_1) \right)^{n+1}   e^{- i \omega_1 (t_3 - t_1)} \nn \\ 
&  \qquad \times  \int  \frac{d\omega_2}{2\pi}   \left( G^*(\omega_2) \right)^{n+1}  e^{- i \omega_2 (t_4 - t_2)}   \,.
\label{masslessOTOC}
\end{align}
To check if our prescription is consistent, it is useful to evaluate these with the free propagator, rather than the dressed one as is done in \cite{Michel:2016kwn}.  Setting 
\begin{align}
G(\omega) = \frac{i}{\omega + i \epsilon},
\end{align}
we have 
\begin{align}
\int  \frac{d\omega}{2\pi} \left( G(\omega) \right)^{n+1} e^{- i \omega t} =  \frac{t^n}{n!} \theta(t) \,, \quad
\int  \frac{d\omega}{2\pi} \left( G^*(\omega) \right)^{n+1} e^{+ i \omega t} =  \frac{t^n}{n!} \theta(t) \,,
\end{align}
then, the time-ordered one eq.~\eqref{zeromasstimeorder} gives 
\begin{align}
&N {G}^{(n)}_4  (t_1, t_2, t_3, t_4)  =  \left( - \frac{\nu_T^2}{2} \right)^n\frac{ \left( t_3 - t_1\right)^n \left( t_4 - t_2\right)^n   }{( n!)^2}  \,   \theta(t_3 - t_1) \, \theta(t_4 - t_2)   \,.
\end{align}
On the other hand, the out-of-time-ordered one eq.~\eqref{masslessOTOC} gives 
\begin{align}
&N {G}^{(n)}_4  (t_1, t_2, t_3, t_4)  =   \left(  \frac{\nu_T^2}{2} \right)^n \frac{ \left( t_3 - t_1\right)^n \left( t_2 - t_4\right)^n   }{( n!)^2}  \, \theta(t_3 - t_1) \, \theta(t_2 - t_4)   \,.
\end{align}
Thus, after summing over all the rungs, for the time-ordered one we have 
\begin{align}
\hspace{-1mm}N {G}_4  (t_1, t_2, t_3, t_4)  &=   \theta(t_3 - t_1) \, \theta(t_4 - t_2)\,  \sum_{n=1}^\infty   \left( - \frac{\nu_T^2}{2} \right)^n \frac{ \left( t_3 - t_1\right)^n \left( t_4 - t_2\right)^n   }{( n!)^2}  \nn \\
& = \theta(t_3 - t_1) \, \theta(t_4 - t_2)\, \left( {J_0\left(  \sqrt{2 \nu^2_T \left( t_3 - t_1 \right) \left(t_4 -  t_2 \right) }  \right) - 1 } \right) \,,
\label{freeG4timeorder}
\end{align}
and for the out-of-time-ordered one, we have 
\begin{align}
N {G}_4  (t_1, t_2, t_3, t_4)  &=   \theta(t_3 - t_1) \, \theta(t_2 - t_4)\,  \sum_{n=1}^\infty   \left( \frac{\nu_T^2}{2} \right)^n \frac{ \left( t_3 - t_1\right)^n \left( t_4 - t_2\right)^n   }{( n!)^2}  \nn \\
& = \theta(t_3 - t_1) \, \theta(t_2 - t_4)\, \left( {I_0\left(  \sqrt{2 \nu^2_T \left( t_3 - t_1 \right) \left(t_4 -  t_2 \right) }  \right) - 1 }\right) \,.
\label{freeG4outoftimeorder}
\end{align}
One can see that by dropping the theta function $\theta(t_4-t_2)$, and by making the analytic continuation as $t_4 - t_2 \to t_2 - t_4$, eq.~\eqref{freeG4timeorder} becomes eq.~\eqref{freeG4outoftimeorder} \cite{Michel:2016kwn} and thus it is consistent wtih the analytic continuation of time.

\subsection{Analytic continuation for $m > 0$ OTOCs}

Now we consider general nonzero mass $m > 0$ cases. For the out-of-time-ordered correlators, we can conduct the $p$ integral in eq.~\eqref{prescription} explicitly. 
 
Let us start with the simplest example, $ {G}^{(1)}_4$, a single rung one for the time-ordered case. In momentum space, it is given by 
\begin{align}  
N {G}^{(1)}_4  = - \lambda \int \frac{dp_1}{2\pi} G(\omega_1) G(\omega_1 - p_1) G(\omega_2) G(\omega_2+p_1) K(p_1) \nn \\
\times \left( 2 \pi \right)^2 \delta(w_1- \omega_3 - p_1 ) \delta(w_2 - \omega_4 + p_1 )  \,,
\end{align}
where $K(p_1)$ is given by eq.~\eqref{ktherm}.  
By the prescription of eq.~\eqref{prescription}, for the out-of-time-ordered one, this becomes 
\begin{align} \label{eq:singleOTOC}
N {G}^{(1)}_4  = \lambda \int \frac{dp_1}{2\pi} G(\omega_1) G(\omega_1 - p_1) G^*(\omega_2) G^*(\omega_2+p_1) K(p_1) \nn \\
\times \left( 2 \pi \right)^2 \delta(w_1- \omega_3 - p_1 ) \delta(w_2 - \omega_4 + p_1 )  \,,
\end{align}
Now we can conduct $p_1$ integration. Since $G(\omega)$ ($G^*(\omega)$) has no singularities in the upper (lower) half plane, one can close the contour by going to the lower half plane for $p_1$ and then it picks up only the pole of $K(p_1)$ and then it yields 
\begin{align} \label{eq:singleOTOC2}
 \hspace{-5mm}N {G}^{(1)}_4  &= \frac{\lambda}{2 m \left(1-y \right)} \Bigl[  G(\omega_1) G(\omega_1 - m) G^*(\omega_2) G^*(\omega_2+m)   \left( 2 \pi \right)^2 \delta(w_1- \omega_3 - m ) \delta(w_2 - \omega_4 + m ) \nn \\
 &+ y \, G(\omega_1) G(\omega_1 + m) G^*(\omega_2) G^*(\omega_2 - m)   \left( 2 \pi \right)^2 \delta(w_1- \omega_3 +m ) \delta(w_2 - \omega_4 -m ) \Bigr]  \nn \\
& =  \frac{\lambda}{2 m \left(1-y \right)} \sum_{n_1 = \pm 1}  \Bigl( y^{(1-n_1)/2} \, G(\omega_1) G(\omega_1 - n_1 m) G^*(\omega_2) G^*(\omega_2 + n_1 m)  \nn \\
&\qquad \qquad \qquad \qquad \times \left( 2 \pi \right)^2 \delta(w_1- \omega_3 - n_1 m ) \delta(w_2 - \omega_4 + n_1  m) \Bigr) \,.
\end{align}
Similarly, we can obtain the out-of-time-ordered correlators with $n$ rung as   
\begin{align}
& \hspace{-5mm}N {G}^{(n)}_4  = \left( \frac{\lambda}{2 m \left(1-y \right)} \right)^n \sum_{n_1 , n_2 , \cdots \,, n_n } y^{(n- \sum_{i=1}^n n_i)/2} \, G(\omega_1) G(\omega_1 - n_1 m)   G(\omega_1 - (n_1 + n_2) m)  \nn \\
& \cdots  G(\omega_1 - \sum_{i=1}^n n_i m) \,\times G^*(\omega_2) G^*(\omega_2 + n_n m) G^*(\omega_2 + (n_n + n_{n-1}) m) \cdots   G^*(\omega_2 +  \sum_{i=1}^n n_i m) \nn \\
&\qquad \qquad \qquad \qquad \times \left( 2 \pi \right)^2 \delta(w_1- \omega_3 - \sum_{i=1}^n n_i m ) \delta(w_2 - \omega_4 + \sum_{i=1}^n n_i m) \,.
\end{align}

Then we can obtain the position space expressions for the $n$ rung out-of-time-ordered correlators from \eqref{n-thGrealtime} as 
\begin{align}
& \hspace{-5mm}N {G}^{(n)}_4 (t_1, t_2, t_3, t_4) = \int  \frac{d\omega_1}{2\pi}   \frac{d\omega_2}{2\pi}   \frac{d\omega_3}{2\pi}   \frac{d\omega_4}{2\pi} N {G}^{(n)}_4  (\omega_1, \omega_2, \omega_3, \omega_4) e^{i\omega_1t_1+i\omega_2t_2-i\omega_3 t_3-i \omega_4 t_4} \nn \\
& =   \left( \frac{\lambda}{2 m \left(1-y \right)} \right)^n \sum_{n_1 , n_2 , \cdots \,, n_n } y^{(n- \sum_{i=1}^n n_i)/2}  \, e^{+ i  \sum_i n_i m (t_3 - t_4)} \nn \\
& \int  \frac{d\omega_1}{2\pi}  G(\omega_1) G(\omega_1 - n_1 m)   G(\omega_1 - (n_1 + n_2) m)   \cdots  G(\omega_1 - \sum_{i=1}^n n_i m) e^{ -i \omega_1(t_3 - t_1)} \, \nn \\
& \times \int \frac{d\omega_2}{2\pi} G^*(\omega_2) G^*(\omega_2 + n_n m) G^*(\omega_2 + (n_n + n_{n-1}) m) \cdots   G^*(\omega_2 +  \sum_{i=1}^n n_i m) e^{-i \omega_2  (t_4 - t_2) }  \nn \\
& \quad \propto \theta(t_3 - t_1) \theta(t_2 - t_4) \,.
\end{align}
For OTOCs, let us set 
\be
t_1 = t_2=0 \,, \quad t_3 = - t_4 = t \,.
\ee
Then by defining  $G(n_1, n_2, \cdots , n_n, t)$ as 
\be
\label{Gnnt}
G(n_1, n_2, \cdots , n_n, t) := \int  \frac{d\omega}{2\pi}  G(\omega) G(\omega - n_1 m)   G(\omega - (n_1 + n_2) m)   \cdots  G(\omega - \sum_{i=1}^n n_i m) e^{ -i \omega t },
\ee
the OTOCs becomes 
\begin{align}\label{Gnt}
 \hspace{-5mm}N {G}^{(n)}_4 (0, 0, t, - t) & = \left( \frac{\lambda}{2 m \left(1-y \right)} \right)^n \sum_{n_1 , n_2 , \cdots \,, n_n } \Bigl[ y^{(n- \sum_{i=1}^n n_i)/2}  \, e^{+ 2 i  \sum_i n_i m \, t} \nn\\
&\qquad \qquad \times G(n_1, n_2, \cdots , n_n, t) G^*(-n_n, -n_{n-1}, \cdots , -n_1, t) \Bigr] \,.
\end{align}
Thus, the final expression for OTOCs is obtained by summing over all rungs as 
\begin{align}
\label{finalOTOCs}
N {G}_4 (0, 0, t, - t)   &= \sum_{n=1}^\infty \left( \frac{\nu_T}{2 } \right)^{2n} \sum_{n_1 , n_2 , \cdots \,, n_n } \Bigl[ y^{(n- \sum_{i=1}^n n_i)/2}  \, e^{+ 2 i  \sum_i n_i m \, t} \ G(n_1, n_2, \cdots , n_n, t) \nn \\
& \qquad \qquad \qquad \qquad \qquad \qquad \times G^*(-n_n, -n_{n-1}, \cdots , -n_1, t) \Bigr] \,,
\end{align}
where $n_i$ takes $+1$ or $-1$, and $y$ and $\nu_T$ are given by eq.~\eqref{defnuT} and \eqref{ktherm} respectively.

Sometimes it is more convenient to evaluate OTOCs where we analytically continue 
\begin{align}
t_2 \to t_2 - \frac{i \beta}{2} \,, \quad t_4 \to t_4 - \frac{i \beta}{2}  \,, 
\end{align} 
rather than Eq.~\eqref{finalOTOCs}.  
In this case, the propagator $K(p)$ in eq.~\eqref{ktherm} is replaced as 
\begin{align}
K(p) \to K(p) e^{- \frac{\beta p}{2}}.
\end{align}
Since for OTOCs, we pick up the pole at $\omega = m$ in the first time of eq.~\eqref{ktherm}, and $\omega = - m $ in the second term of eq.~\eqref{ktherm},  this essentially replace the $K(p)$  in eq.~\eqref{ktherm} as 
\begin{align}
K(p) \to  \frac{i}{1-y}\left(\frac{\sqrt{y}}{p^2-m^2 +i \epsilon} - \frac{\sqrt{y}}{p^2-m^2-i\epsilon}\right)~.
\end{align}
and this simply ends up replacing the $y$-dependent factors into $n_i$ independent one and we obtain 
\begin{align}
\label{finalOTOCsbeta2}
&N {G}_4 (0, -i \frac{\beta}{2}, t, - t-i \frac{\beta}{2})   \nn \\
&= \sum_{n=1}^\infty \left( \frac{ y^{1/4} \, \nu_T}{2 } \right)^{2n} \sum_{n_1 , n_2 , \cdots \,, n_n } \Bigl[   \, e^{+ 2 i  \sum_i n_i m \, t} \ G(n_1, n_2, \cdots , n_n, t) 
G^*(-n_n, -n_{n-1}, \cdots , -n_1, t) \Bigr] \,. 
\end{align}

We will evaluate both \eqref{finalOTOCs} and \eqref{finalOTOCsbeta2} numerically for various parameters in the IP matrix model in the next section. 

However, before we proceed, we will comment on some important properties of $N {G}_4 (0, 0, t, - t) $ in eq.~\eqref{finalOTOCs} and $N {G}_4 (0, -i \frac{\beta}{2}, t, - t-i \frac{\beta}{2}) $ in  eq.~\eqref{finalOTOCsbeta2}. 
\begin{enumerate} 
\item Both $N {G}_4 (0, 0, t, - t) $ and $N {G}_4 (0, -i \frac{\beta}{2}, t, - t-i \frac{\beta}{2})$ depend on the three dimensionful parameters, the adjoint mass $m > 0$, temperature $T$, and $\nu_T$ (or equivalently 't Hooft coupling $\lambda/m$). Thus its time dependence is generically quite complicated.  
\item At the infinite temperature limit, $y=1$, $\beta =0$, eq.~\eqref{finalOTOCsbeta2} becomes identical to eq.~\eqref{finalOTOCs}. 
\item The phase factor $e^{+ 2 i  \sum_i n_i m \, t}$ gives oscillation, and its periodicity is proportional to $1/m$.\footnote{Periodicity from this phase factor is $T =\pi/m$. However, numerically we will find periodicity is $T \approx 2 \pi/m$.} To check if the magnitudes of OTOCs $N {G}_4 (0, 0, t, - t) $ and $N {G}_4 (0, -i \frac{\beta}{2}, t, - t-i \frac{\beta}{2}) $ grow, we need to take into account this oscillation in time.  
\item In the massless limit $m\to 0$ where $T$ is fixed, then $y \to 1$ and all the $n_i$ dependence on $G(n_1, n_2, \cdots , n_n, t)$ disappears as
\begin{align}
\label{Gnntmassless}
\lim_{m \to 0}G(n_1, n_2, \cdots , n_n, t) &=  \int  \frac{d\omega}{2\pi}  \left( G(\omega) \right)^{n+1} e^{ -i \omega t }
\end{align}
and then both eq.~\eqref{finalOTOCs} and \eqref{finalOTOCsbeta2} reduce to 
\begin{align}
\label{finalmasslessOTOCs}
N {G}_4 (0, 0, t, - t)   &= N {G}_4 (0, -i \frac{\beta}{2}, t, - t-i \frac{\beta}{2})  \nn \\
&=  \sum_{n=1}^\infty \left( \frac{\nu_T}{2 } \right)^{2n} 2^n \,  \left| \int  \frac{d\omega}{2\pi}  \left( G(\omega) \right)^{n+1} e^{ -i \omega t } \right|^2
\end{align}
\item We can also take the double-scaling limit for massless cases as 
\be
m \to 0 \,, \quad T \to 0 \,, \quad y  = \mbox{fixed}
\ee
Then from eq.~\eqref{finalOTOCsbeta2}, we obtain 
\begin{align}
\label{finalOTOCsbeta3}
&N {G}_4 (0, -i \frac{\beta}{2}, t, - t-i \frac{\beta}{2}) = \sum_{n=1}^\infty \left( \frac{ y^{1/4} \, \nu_T}{2 } \right)^{2n} 2^n \,  \left| \int  \frac{d\omega}{2\pi}  \left( G(\omega) \right)^{n+1} e^{ -i \omega t } \right|^2\end{align}
where $G(\omega)$ is given by \eqref{doublezeroGomega}. Note that in this case, the radius of the Wigner semi-circle is given by $\sqrt{1+y} \, \nu_T$, on the other hand, the coefficient in the $2n$-th power of \eqref{finalOTOCsbeta3} is proportional to $\sqrt{2} y^{1/4} \nu_T$. Unless $y=1$, we have two different scale $\sqrt{1+y} \, \nu_T > \sqrt{2} y^{1/4} \nu_T$. 
\end{enumerate}
In the next section, 
we will evaluate both eq.~\eqref{finalOTOCs} and \eqref{finalOTOCsbeta2} numerically for various parameters

\section{Numerical analysis for OTOCs in various parameters}
\label{mainnumerics}

In this section, we show numerical computations of the OTOCs with our prescription. 
To evaluate them numerically, we consider a finite sum of $n$ up to $n_{max}$ in eq.~\eqref{finalOTOCs} and define
\begin{align}
\label{finalOTOCsnmax}
N {G}^{n_{max}}_4 (0, 0, t, - t)   &:= \sum_{n=1}^{n_{max}} \left( \frac{\nu_T}{2 } \right)^{2n} \sum_{n_1 , n_2 , \cdots \,, n_n } \Bigl[ y^{(n- \sum_{i=1}^n n_i)/2}  \, e^{+ 2 i  \sum_i n_i m \, t} \ G(n_1, n_2, \cdots , n_n, t) \nn \\
& \qquad \qquad \qquad \qquad \qquad \qquad \times G^*(-n_n, -n_{n-1}, \cdots , -n_1, t) \Bigr] \,. 
\end{align}
Similarly from eq.~\eqref{finalOTOCsbeta2}, we also define 
\begin{align}
\label{finalOTOCsnmaxv2}
N {G}^{n_{max}}_4 (0, -i\frac{\beta}{2}, t, - t -i\frac{\beta}{2})    &:= \sum_{n=1}^{n_{max}} \left( \frac{y^{1/4}\nu_T}{2 } \right)^{2n} \sum_{n_1 , n_2 , \cdots \,, n_n } \Bigl[ \, e^{+ 2 i  \sum_i n_i m \, t} \ G(n_1, n_2, \cdots , n_n, t) \nn \\
& \qquad \qquad \qquad \qquad \qquad \qquad \times G^*(-n_n, -n_{n-1}, \cdots , -n_1, t) \Bigr] \,.
\end{align}
Note that the two expressions are identical if $y=1$.

\subsection{Double-scaling limit $m\to0$, $T\to0$, $y=\text{fixed}$}
In the double-scaling limit $m\to0$, $T\to0$, $y=\text{fixed}$, the spectrum is given by eq.~\eqref{doublezeroGomega}.
In this limit, $G(n_1, n_2, \cdots , n_n, t)$, defined by eq.~\eqref{Gnnt} does not depend on $n_i$ as \eqref{Gnntmassless}. With the spectrum (\ref{doublezeroGomega}), the OTOCs (\ref{finalOTOCsnmax}) and (\ref{finalOTOCsnmaxv2}) are simplified as
\begin{align}
\label{finalmasslessOTOCsdoublescalingnmax}
&N {G}^{n_{max}}_4 (0, 0, t, - t)   =  \sum_{n=1}^{n_{max}} \left( \frac{\nu_T^2(1+y)}{4} \right)^{n}  \,  \left| \int  \frac{d\omega}{2\pi}  \left( G(\omega) \right)^{n+1} e^{ -i \omega t } \right|^2 , \\
&N {G}^{n_{max}}_4 (0, -i\frac{\beta}{2}, t, - t -i\frac{\beta}{2})   =  \sum_{n=1}^{n_{max}} \left( \frac{y^{1/2}\nu_T^2 }{2} \right)^{n}  \,  \left| \int  \frac{d\omega}{2\pi}  \left( G(\omega) \right)^{n+1} e^{ -i \omega t } \right|^2.\label{finalmasslessOTOCsdoublescalingnmax2}
\end{align}

\begin{figure}[t]
\centering
     \begin{subfigure}[b]{0.6\textwidth}
         \centering
         \includegraphics[width=\textwidth]{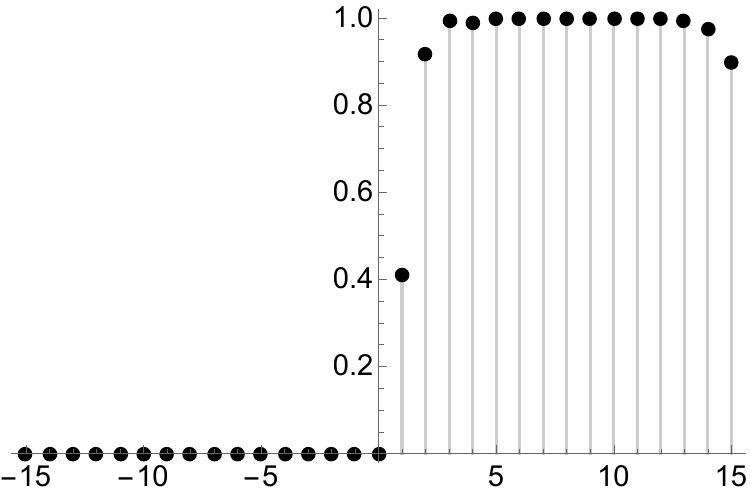}
       \put(5,15){$t$}
    \put(-190,185){$\text{Re}[N {G}^{n_{max}}_4 (0, 0, t, - t)]$}
\label{fig:OTOCWigner1}
\caption{OTOC given by eq.~(\ref{finalmasslessOTOCsdoublescalingnmax}).}
\label{fig:OTOCWigner(a)}
      \vspace{10mm} 
     \end{subfigure}
      \quad\quad\quad   
     \begin{subfigure}[b]{0.6\textwidth}
         \centering
         \includegraphics[width=\textwidth]{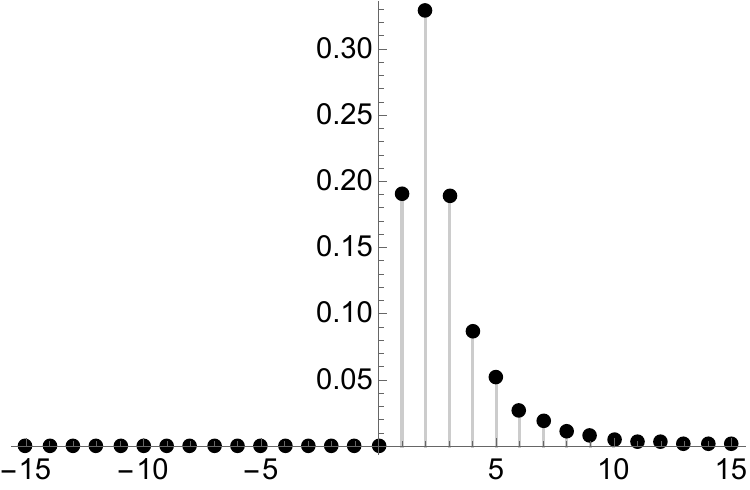}
          \put(5,15){$t$}
 \put(-190,185){$\text{Re}[N {G}^{n_{max}}_4 (0, -i\frac{\beta}{2}, t, - t -i\frac{\beta}{2})]$}
\label{fig:OTOCWigner2}
\caption{OTOC given by eq.~(\ref{finalmasslessOTOCsdoublescalingnmax2}).}
\label{fig:OTOCWigner(b)}
     \end{subfigure}
             \caption{OTOCs given by eq.~(\ref{finalmasslessOTOCsdoublescalingnmax}) and eq.~(\ref{finalmasslessOTOCsdoublescalingnmax2}) as a function of time $t$. We sum rungs up to $n_{max}=20$ with $\nu_T^2(1+y)=2$ and $y^{1/2}\nu_T^2=1/2$.}
        \label{fig:OTOCWigner}
\end{figure}

Figure \ref{fig:OTOCWigner} shows both $N {G}^{n_{max}}_4 (0, 0, t, - t)$ and $N {G}^{n_{max}}_4 (0, -i\frac{\beta}{2}, t, - t -i\frac{\beta}{2})$ given by \eqref{finalmasslessOTOCsdoublescalingnmax} and \eqref{finalmasslessOTOCsdoublescalingnmax2} as a function of time $t$, with $n_{max}=20$, $\nu_T^2(1+y)=2$ and $y^{1/2}\nu_T^2=1/2$. In both figures, the OTOCs are zero when $t<0$ since we use the retarded Green function spectrum $G(\omega)$. Figure \ref{fig:OTOCWigner(a)} shows that the OTOC (\ref{finalmasslessOTOCsdoublescalingnmax}) saturates to 1. 
Note that this result is consistent with the result obtained in \cite{Michel:2016kwn}.\footnote{Since the radius of Wigner semi-circle is $\sqrt{1+y} \, \nu_T$ and since the $2n$-th power of the coefficients is proportional to $\sqrt{1+y} \, \nu_T $, both are the same scale and thus there is only one scale $\nu$. Note that there are notation differences between us and  \cite{Michel:2016kwn}; $\nu$ in \cite{Michel:2016kwn} corresponds to our $\nu_T$.}  The decrease around $t \sim 15$ is due to the finite sum of $n$ up to $n_{max}=20$. By increasing $n_{max}$, we can check the value stays around $1$ even after $ t \sim 15$. 

In contrast to the OTOC (\ref{finalmasslessOTOCsdoublescalingnmax}), Figure \ref{fig:OTOCWigner(b)} shows that the OTOC (\ref{finalmasslessOTOCsdoublescalingnmax2}) does not saturate to 1. This is because the OTOC (\ref{finalmasslessOTOCsdoublescalingnmax2}) depends on two combinations of the parameters $\nu_T^2(1+y)/2$ and $y^{1/2}\nu_T^2$. Since the factor $\left( \frac{y^{1/2}\nu_T^2 }{2} \right)^{n}$ in (\ref{finalmasslessOTOCsdoublescalingnmax2}) is smaller than the factor $\left( \frac{\nu_T^2(1+y)}{4} \right)^{n}$ in (\ref{finalmasslessOTOCsdoublescalingnmax}), the OTOC  (\ref{finalmasslessOTOCsdoublescalingnmax2}) does not saturate to 1 in our numerical computation, but rather it decays. Note that 
\be
y^{1/2}\nu_T^2\le \nu_T^2(1+y)/2 \quad \mbox{as long as $m/T\ge0$}  \,,
\ee 
where $y=e^{-m/T}$.

\newpage

\subsection{Nonzero mass case $m > 0$}
\begin{figure}[H]
\centering{
       \begin{subfigure}[b]{0.73\textwidth}
         \centering
         \includegraphics[width=\textwidth]{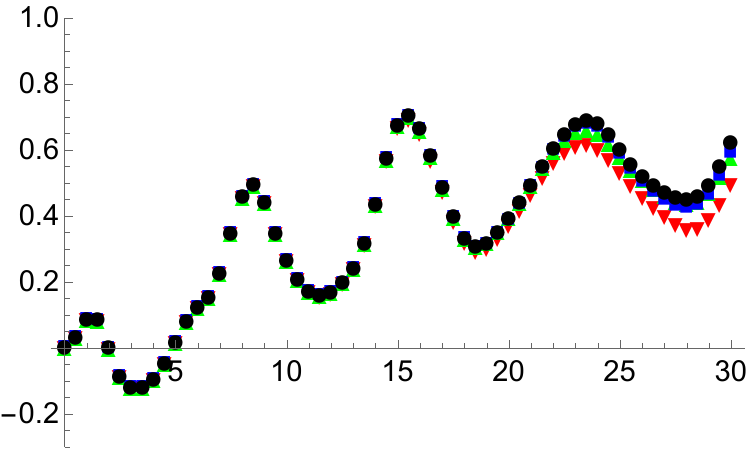}
       \put(5,40){$t$}
    \put(-325,195){$\text{Re}[N {G}^{n_{max}}_4 (0, -i\frac{\beta}{2}, t, - t -i\frac{\beta}{2})]$}
       \caption{$m=0.8, \;\nu_T=1, \;y=0.1.$}\label{fig:OTOCm08v1y01}
     \vspace{10mm} 
     \end{subfigure}
        \begin{subfigure}[b]{0.73\textwidth}
         \centering
         \includegraphics[width=\textwidth]{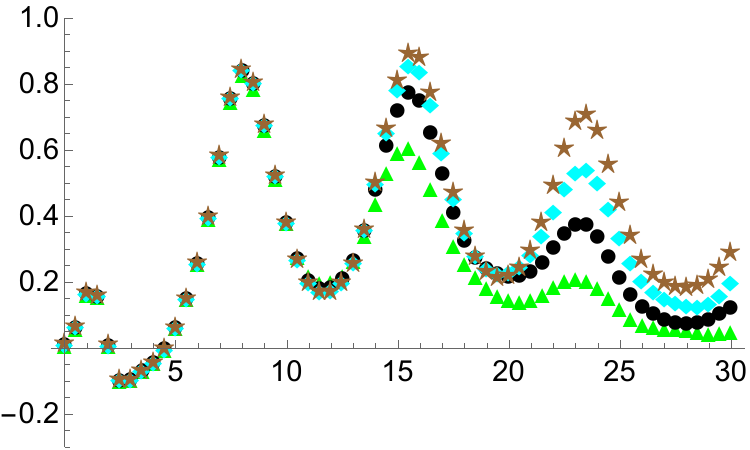}
   \put(5,40){$t$}
    \put(-325,195){$\text{Re}[N {G}^{n_{max}}_4 (0, -i\frac{\beta}{2}, t, - t -i\frac{\beta}{2})]$}
         \caption{$m=0.8, \;\nu_T=1, \;y=0.4$.}\label{fig:OTOCm08v1y04}
    \vspace{5mm} 
     \end{subfigure}
      \begin{subfigure}[r]{0.35\textwidth}
         \includegraphics[width=1.45\textwidth]{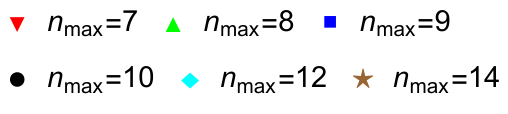}
         \caption*{}
     \end{subfigure}
   \caption{Numerical plots of the OTOCs $\text{Re}[N {G}^{n_{max}}_4 (0, -i\frac{\beta}{2}, t, - t -i\frac{\beta}{2})]$ for $m=0.8$, $\nu_T=1$, (a) $y=0.1$ and (b) $ y=0.4$. }
        \label{fig:OTOCm08v1}
        }
\end{figure}

\begin{figure}[H]
 \centering{
       \begin{subfigure}[b]{0.73\textwidth}
         \centering
         \includegraphics[width=\textwidth]{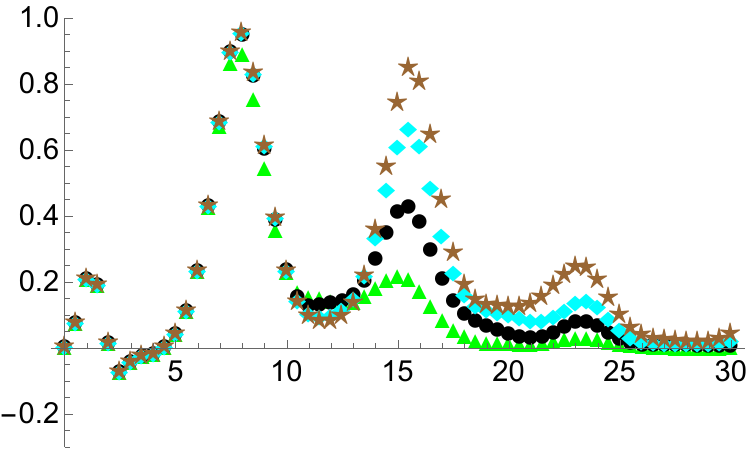}
       \put(5,40){$t$}
    \put(-325,195){$\text{Re}[N {G}^{n_{max}}_4 (0, -i\frac{\beta}{2}, t, - t -i\frac{\beta}{2})]$}
       \caption{$m=0.8$, $\nu_T=1$, $y=0.7$.}\label{fig:OTOCm08v1y07}
            \vspace{10mm} 
     \end{subfigure}
        \begin{subfigure}[b]{0.73\textwidth}
         \centering
         \includegraphics[width=\textwidth]{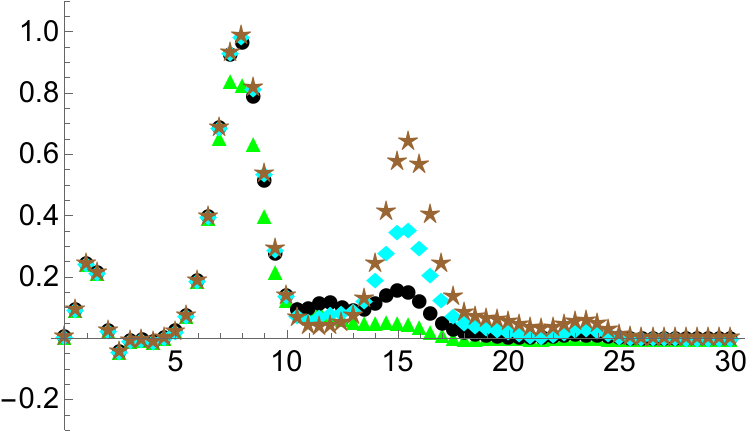}
   \put(5,40){$t$}
    \put(-325,195){$\text{Re}[N {G}^{n_{max}}_4 (0, -i\frac{\beta}{2}, t, - t -i\frac{\beta}{2})]$}
         \caption{$m=0.8$, $\nu_T=1$, $y=1$.}\label{fig:OTOCm08v1y1}
              \vspace{5mm} 
     \end{subfigure}
          \quad  \quad  \quad\quad\quad\quad\quad
      \begin{subfigure}[r]{0.35\textwidth}
         \centering
         \includegraphics[width=1.45\textwidth]{OTOCLegend.pdf}
         \caption*{}
     \end{subfigure}
                                 \caption{Numerical plots of the OTOCs $\text{Re}[N {G}^{n_{max}}_4 (0, -i\frac{\beta}{2}, t, - t -i\frac{\beta}{2})]$ for $m=0.8, \;\nu_T=1$, (a) $y=0.7$ and  (b) $y=1$. }
        \label{fig:OTOCm08v2}
}
\end{figure}

\begin{figure}[H]
\centering{
       \begin{subfigure}[b]{0.73\textwidth}
         \centering
         \includegraphics[width=\textwidth]{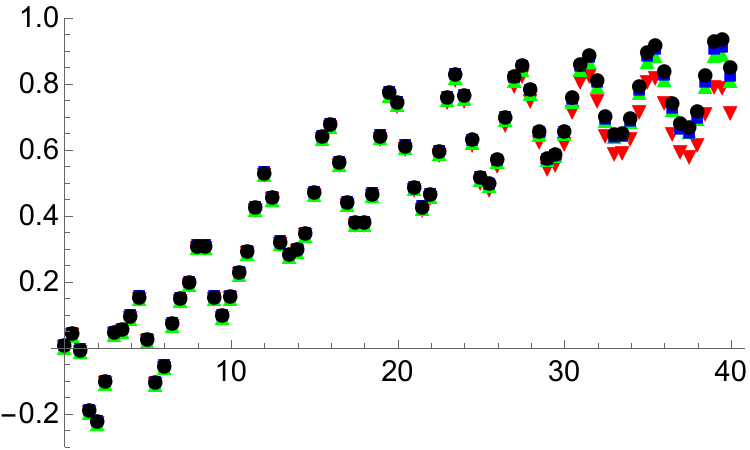}
       \put(5,40){$t$}
    \put(-325,195){$\text{Re}[N {G}^{n_{max}}_4 (0, 0, t, - t)]$}
       \caption{$m=1.6, \;\nu_T=1, \;y=0.1$.}\label{fig:OTOCm16v1y01}
            \vspace{10mm} 
     \end{subfigure}
        \begin{subfigure}[b]{0.73\textwidth}
         \centering
         \includegraphics[width=\textwidth]{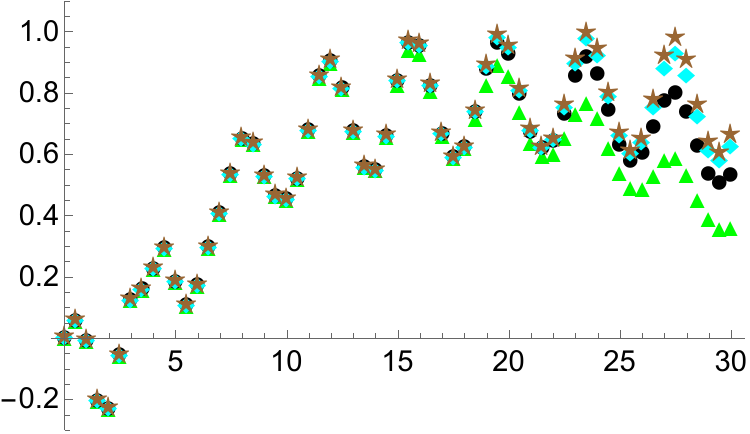}
   \put(5,40){$t$}
    \put(-325,195){$\text{Re}[N {G}^{n_{max}}_4 (0, 0, t, - t)]$}
         \caption{$m=1.6, \;\nu_T=1, \;y=0.4$.}\label{fig:OTOCm16v1y04}
              \vspace{5mm} 
     \end{subfigure}
          \quad  \quad  \quad\quad\quad\quad\quad
      \begin{subfigure}[r]{0.35\textwidth}
         \centering
         \includegraphics[width=1.45\textwidth]{OTOCLegend.pdf}
         \caption*{}
     \end{subfigure}
                                 \caption{Numerical plots of the OTOCs $\text{Re}[N {G}^{n_{max}}_4 (0, 0, t, - t)]$ for $m=1.6, \;\nu_T=1$, (a) $y=0.1$ and (b) $y= 0.4$. }
        \label{fig:OTOCm16v1}
        }
\end{figure}

\begin{figure}[H]
 \centering{
       \begin{subfigure}[b]{0.73\textwidth}
         \centering
         \includegraphics[width=\textwidth]{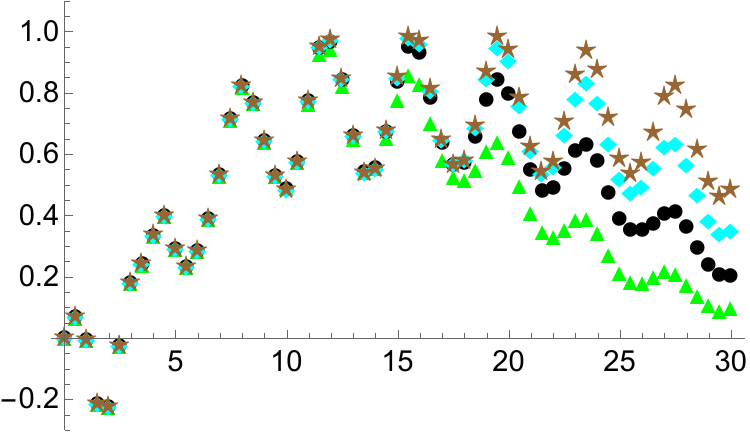}
       \put(5,40){$t$}
    \put(-325,195){$\text{Re}[N {G}^{n_{max}}_4 (0, 0, t, - t)]$}
       \caption{$m=1.6, \;\nu_T=1, \;y=0.7$.}\label{fig:OTOCm16v1y07}
            \vspace{10mm} 
     \end{subfigure}
        \begin{subfigure}[b]{0.73\textwidth}
         \centering
         \includegraphics[width=\textwidth]{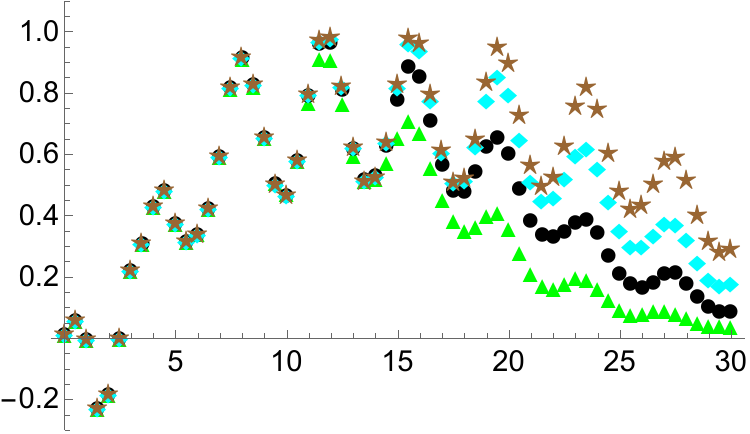}
   \put(5,40){$t$}
    \put(-325,195){$\text{Re}[N {G}^{n_{max}}_4 (0, 0, t, - t)]$}
         \caption{$m=1.6, \;\nu_T=1, \;y=1$.}\label{fig:OTOCm16v1y1}
              \vspace{5mm} 
     \end{subfigure}
          \quad  \quad  \quad\quad\quad\quad\quad
      \begin{subfigure}[r]{0.35\textwidth}
         \centering
         \includegraphics[width=1.45\textwidth]{OTOCLegend.pdf}
         \caption*{}
     \end{subfigure}
                                 \caption{Numerical plots of the OTOCs $\text{Re}[N {G}^{n_{max}}_4 (0, 0, t, - t)]$ for $m=1.6, \;\nu_T=1$, (a) $y=0.7$ and (b) $y=1$. }
        \label{fig:OTOCm16v2}
        }
\end{figure}

\begin{figure}[H]
\centering{
       \begin{subfigure}[b]{0.73\textwidth}
         \centering
         \includegraphics[width=\textwidth]{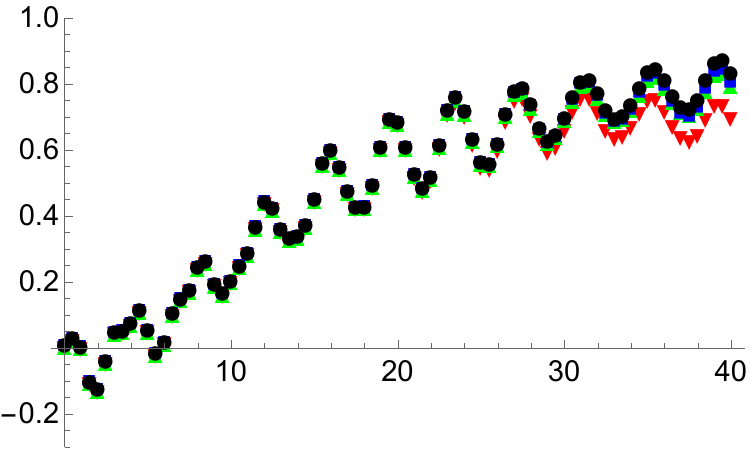}
   \put(5,40){$t$}
    \put(-325,195){$\text{Re}[N {G}^{n_{max}}_4 (0, -i\frac{\beta}{2}, t, - t -i\frac{\beta}{2})]$}
         \caption{$m=1.6, \;\nu_T=1, \;y=0.1$.}\label{fig:OTOCm16v1y01DC}
              \vspace{10mm} 
     \end{subfigure}
        \begin{subfigure}[b]{0.73\textwidth}
         \centering
          \includegraphics[width=\textwidth]{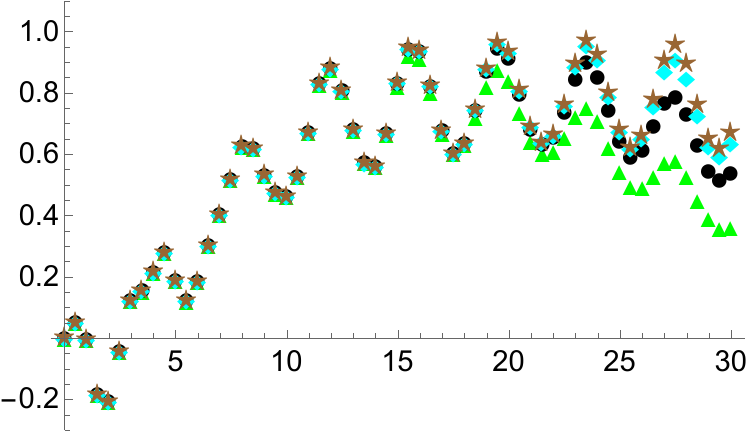}
   \put(5,40){$t$}
    \put(-325,195){$\text{Re}[N {G}^{n_{max}}_4 (0, -i\frac{\beta}{2}, t, - t -i\frac{\beta}{2})]$}
         \caption{$m=1.6, \;\nu_T=1, \;y=0.4$.}\label{fig:OTOCm16v1y04DC}
              \vspace{5mm} 
     \end{subfigure}
              \quad  \quad  \quad\quad\quad\quad\quad
           \begin{subfigure}[r]{0.35\textwidth}
         \centering
\includegraphics[width=1.45\textwidth]{OTOCLegend.pdf}
         \caption*{}
     \end{subfigure}
     \caption{Numerical plots of the OTOCs $\text{Re}[N {G}^{n_{max}}_4 (0, -i\frac{\beta}{2}, t, - t -i\frac{\beta}{2})]$ for $m=1.6, \;\nu_T=1$, (a) $y=0.1$ and (b) $y= 0.4$.  }
        \label{fig:OTOCm16v3}
        }
\end{figure}

\begin{figure}[H]
 \centering{
       \begin{subfigure}[b]{0.73\textwidth}
         \centering
         \includegraphics[width=\textwidth]{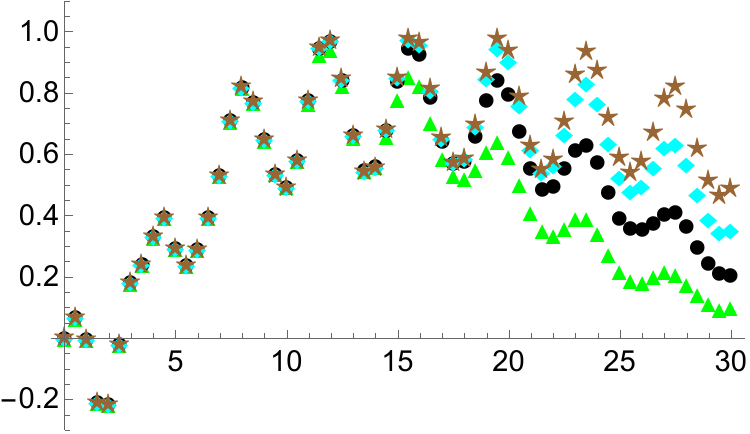}
   \put(5,40){$t$}
    \put(-325,195){$\text{Re}[N {G}^{n_{max}}_4 (0, -i\frac{\beta}{2}, t, - t -i\frac{\beta}{2})]$}
         \caption{$m=1.6, \;\nu_T=1, \;y=0.7$.}\label{fig:OTOCm16v1y01DC}
     \end{subfigure}
              \quad  \quad  \quad\quad\quad\quad\quad
           \begin{subfigure}[r]{0.35\textwidth}
         \centering
\includegraphics[width=1.45\textwidth]{OTOCLegend.pdf}
         \caption*{}
     \end{subfigure}
     \caption{Numerical plots of the OTOCs $\text{Re}[N {G}^{n_{max}}_4 (0, -i\frac{\beta}{2}, t, - t -i\frac{\beta}{2})]$ for $m=1.6, \;\nu_T=1, \;y=0.7$.  }
        \label{fig:OTOCm16v4}
        }
\end{figure}

When mass $m$ is nonzero, the spectrum changes its shape depending on temperature $T$ as shown in Figure \ref{fig:ReG(w)}. 
By substituting numerical data of $G(\omega)$ for nonzero mass, we can numerically compute $G(n_1, n_2, \cdots , n_n, t)$ defined in eq.~(\ref{Gnnt}), and thus the OTOCs given by eq.~(\ref{finalOTOCsnmax}) and (\ref{finalOTOCsnmaxv2}). Figures \ref{fig:OTOCm08v1} and  \ref{fig:OTOCm08v2} are $\text{Re}[N {G}^{n_{max}}_4 (0, -i\frac{\beta}{2}, t, - t -i\frac{\beta}{2})]$ for $m=0.8$, $y=0.1$, $0.4$, $0.7$, $1$. Figures
\ref{fig:OTOCm16v1} and \ref{fig:OTOCm16v2} are $\text{Re}[N {G}^{n_{max}}_4 (0, 0, t, - t)]$ for $m=1.6$, $y=0.1$, $0.4$, $0.7$, $1$. Figures
\ref{fig:OTOCm16v3} and  \ref{fig:OTOCm16v4} are $\text{Re}[N {G}^{n_{max}}_4 (0, -i\frac{\beta}{2}, t, - t -i\frac{\beta}{2})]$ for $m = 1.6$, $y=0.1$, $0.4$, $0.7$, $1$. 
We set $\nu_T=1$ always. 

From our numerical results, the following properties of the OTOCs can be read.
\begin{itemize}
\vspace{-2mm}
\item Due to the finiteness of $n_{max}$, we trust these figures only at early times. More precisely, we can trust these figures only in the time range where we can see the convergence by increasing $n_{max}$.

\item For large $n$, the effects of $n$ rung contributions in OTOCs are important only at late times. In other words, $N {G}^{n_{max}}_4 (0, 0, t, - t)$ and $N {G}^{n_{max}}_4 (0, -i\frac{\beta}{2}, t, - t -i\frac{\beta}{2})$ at early times do not depend on the value of $n_{max}$. In fact, we see the convergence of the numerical evaluation in our parameter range of $n_{max}$.  

\item At late times, the convergence of summing over $n$-rung contributions becomes weak for our parameter range of $n_{max}$ and at very late times, our numerical  OTOCs (\ref{finalOTOCsnmax}) and (\ref{finalOTOCsnmaxv2}) decay to zero. This is simply because we did not take into account the contributions of $n$ rungs where $ n > n_{\max}$. 

\item 
When $n_{max}$ and mass $m$ are fixed, 
the larger $y$ is, the larger the increase in OTOCs at earlier times.
Similarly, the larger $y$, the faster the decay of OTOCs. 

\item
The OTOCs oscillate due to nonzero mass. We see its periodicity $T$ is $T\approx 2\pi/m$.

\item
The numerical plots of $\text{Re}[N {G}^{n_{max}}_4 (0, -i\frac{\beta}{2}, t, - t -i\frac{\beta}{2})]$ and $\text{Re}[N {G}^{n_{max}}_4 (0, 0, t, - t)]$ are almost identical.


\item 
In Figures \ref{fig:OTOCm08v1} and \ref{fig:OTOCm08v2}, the OTOCs $\text{Re}[N {G}^{n_{max}}_4 (0, -i\frac{\beta}{2}, t, - t -i\frac{\beta}{2})]$ for $m=0.8$, $\nu_T=1$ have a peak at $t \sim15.5$. For $y=0.4$, $0.7$, $1$, even though $n_{max}$ is set to $n_{max}=10$, the numerical values of $\text{Re}[N {G}^{n_{max}}_4 (0, -i\frac{\beta}{2}, t, - t -i\frac{\beta}{2})]$ at $t \sim15.5$ do not converge in our numerical computations. 
%
%
Therefore in Appendix \ref{app2}, to examine the convergence for larger $n_{max}$, we plot the $n_{max}$-dependence of $\text{Re}[N {G}^{n_{max}}_4 (0, -i\frac{\beta}{2}, t, - t -i\frac{\beta}{2})]$ at $t=15.5$ for $m=0.8$, $\nu_T=1$. In our numerical computations, by increasing $n_{max}$ to $n_{max}\sim18$, the convergence at $t=15.5$ becomes better, and $\text{Re}[N {G}^{n_{max}}_4 (0, -i\frac{\beta}{2}, t, - t -i\frac{\beta}{2})]$ at $t=15.5$ seems to approach 1. We also do a similar plot at $t=23.5$ for the next peak.

\item 
In all numerical computations we performed, the OTOCs do not exceed 1. These results suggest that the OTOCs in the IP model with nonzero mass do not grow exponentially,
at least in the parameter spaces we studied.

\end{itemize}

\newpage

\clearpage

\newpage

\clearpage

\newpage

Our numerics show that the OTOCs in the IP model do not grow even for the massive adjoint case. In Appendix A, we show the results of the numerical evaluation of eq.~\eqref{finalmasslessOTOCs} using the Green function $G(\omega)$ for $m =0.8 > 0$ and it grows exponentially. Note that eq.~\eqref{finalmasslessOTOCs} is obtained in the massless limit and thus, using the Green function $G(\omega)$ for $m > 0$ is not the right thing to do. However, this illustrates that the manifest mass dependence between eq.~\eqref{finalOTOCs} and eq.~\eqref{finalmasslessOTOCs} is extremely crucial since in one case it gives exponential growth but in the other case, it does not.

Before we look at the mass dependence on these OTOC formulas eq.~\eqref{finalOTOCs} and eq.~\eqref{finalmasslessOTOCs} more closely, first, let us recall 
the characteristic properties of the spectral density by nonzero mass. 
\begin{itemize}
\item  At zero temperature $y=0$, the spectral density $\text{Re}[G(\omega)]$ for nonzero mass has delta-functional peaks, as shown in Figure \ref{fig:ReG(w)}. Note that the positions of these delta-functional peaks are not related by $\omega \to \omega+ m$. 
\item As we increase the temperature from $T=0$ into small nonzero $T > 0$, the delta-functional peaks become high-peak cuts with narrow widths. Simultaneously, new low-peak cuts appear which is related to high-peak cuts by $\omega \to \omega+ m$ shift. In this way, at a small nonzero temperature $T  > 0$, we have both high-peak cuts and low-peak cuts in the spectral density. See Figure  \ref{fig:ReG(w)} at $y=0.1$, $m=0.8$  
\item As we increase the temperature, the peaks of high-peak cuts become lower, and the peaks of low-peak cuts become higher. See Figure  \ref{fig:ReG(w)} at $y=0.4$, $y=0.7$, $m=0.8$.   
\item At infinite temperature limit, for $m=0.8$ case, the spectral density becomes gapless. On the other hand, for $m=1.6$ case, it has gaps.  The behavior of infinite temperature limit $y=1$ for $m=1.6$ case is similar to low temperature behavior $y=0.1$ for $m=0.8$ case. 
\end{itemize}
Given these facts, one of the manifest differences between eq.~\eqref{finalOTOCs} and eq.~\eqref{finalmasslessOTOCs} is the difference in $G(n_1, n_2, \cdots , n_n, t)$, which is defined in eq.~\eqref{Gnnt}. For nonzero mass case, in $G(n_1, n_2, \cdots , n_n, t)$, the Green function $G(\omega)$ has $m$-shift product behavior for $\omega$ as $G(\omega) G(\omega - n_1 m)   G(\omega - (n_1 + n_2) m)   \cdots  G(\omega - \sum_{i=1}^n n_i m)$. On the other hand, in the massless limit, this simply reduces to the power of $G(\omega)$ as eq.~\eqref{Gnntmassless}.

Suppose that the interval between the peaks of $G(\omega)$ is $m$. Then, the peaks of $G(\omega)$ in $G(\omega) G(\omega - n_1 m)   G(\omega - (n_1 + n_2) m)   \cdots  G(\omega - \sum_{i=1}^n n_i m)$ can resonate with each other, since if there is a high peak at $\omega$, then there is another high peak at $G(\omega \pm m)$, thus, 
since the peak values are multiplied by each other, their values increase as the number of rungs $n$ increases.
In such cases, the resonance causes the growth of eq.~\eqref{Gnnt}, and therefore the OTOCs in \eqref{finalOTOCs} and \eqref{finalOTOCsbeta2} grows as well.

However, the above assumption that the interval between the peaks is $m$ is actually not right. The interval between the peaks is determined by the interval between the poles of $G(\omega)$ at zero temperature, which is generally different from $m$. Thus, the peak resonance in $G(\omega) G(\omega - n_1 m)   G(\omega - (n_1 + n_2) m)   \cdots  G(\omega - \sum_{i=1}^n n_i m)$ does not occur. For example, at low temperatures, even if there is a high peak at $G(\omega)$, the $m$-shift peak appearing at $G(\omega \pm m)$ is a very low peak and thus, the product of $G(\omega)$ in $G(\omega) G(\omega - n_1 m)   G(\omega - (n_1 + n_2) m)   \cdots  G(\omega - \sum_{i=1}^n n_i m)$ does not resonate with each other. This is the mathematical reason why, using the Green function for $m = 0.8$,  
even though eq.~\eqref{finalmasslessOTOCs} grows, eq.~\eqref{finalOTOCs} does not grow at least at low temperatures. 
It is not obvious how large the resulting OTOC will be since the higher peaks become lower and the lower peaks become higher as the temperature increases, but our numerical results suggest that OTOC does not increase much even if the temperature is increased.

Furthermore, there are additional differences between eq.~\eqref{finalOTOCs} and eq.~\eqref{finalmasslessOTOCs} due to nonzero mass, which are the phase factor, and $y$ dependence. These also give suppression effects for OTOCs. For nonzero mass $m > 0$, $y < 1$ gives further suppression effects in OTOCs. So is the phase factor.


\section{Conclusions and discussions}
\label{conclusion} 
In this paper, we study the out-of-time-ordered correlators (OTOCs) in the IP matrix model. 
OTOCs in the IP matrix model for $m=0$, where $m$ is the mass of the adjoint matrix field $X_{ij}$, were studied in \cite{Michel:2016kwn}. 
Since the IP model shows significantly different behaviors between $m =0$ and $m \neq 0$, 
one obvious and missing calculation is the OTOCs calculation in the IP model for $m > 0$ at nonzero temperature, and that's why we study $m > 0$ case OTOCs in this paper.

To invert the time direction for the computation of the OTOCs, we give a prescription for analytic continuation in time, and by that, the poles of the retarded Green functions in the lower half of the complex $\omega$ plane are now reflected in the upper half $\omega$ plane. 
Then we numerically calculate the OTOCs for various parameter ranges.

First, we study the OTOCs in $m \to 0$ limit with $T \neq 0$. In this case, the OTOCs do not grow, which is consistent with the results previously studied in \cite{Michel:2016kwn}.  
We also study the double-scaled limit where $m\to 0$, $T\to 0$ with $y = e^{-m/T}$ = fixed. Even in these double-scaled cases, we find that OTOCs do not grow.

Next, we study various parameter ranges for $m > 0$ cases.  
The results of OTOCs for $m=0.8$ and $m=1.6$ are shown in Figures \ref{fig:OTOCm08v1}, \ref{fig:OTOCm08v2}, \ref{fig:OTOCm16v1}, \ref{fig:OTOCm16v2}, \ref{fig:OTOCm16v3}, \ref{fig:OTOCm16v4}. When $m \neq 0$, unlike the case when $m=0$, 
the infinite series sum for the number of rungs in OTOCs cannot be written in closed form mathematically, therefore we need to add this up numerically sequentially. We have calculated it numerically as far as it converges. The results showed no OTOC growth in the parameter regions we examined.

The sharp reason why the IP model does not show the growth of OTOC is not clear to us. In some sense, it could be that the structure of the IP model is too simple that it does not show the chaos. One possible reason for that is in the IP model, a fundamental field gets thermalized but an adjoint field does not, since the backreaction of a fundamental into an adjoint is suppressed by $1/N$.  In this sense, the IP model shows only {\it partial} thermalization, {\it i.e.,} only a fundamental field gets thermalized but not an adjoint.  As is pointed out in \cite{Michel:2016kwn}, if all of the fields get thermalized, then it is reasonable to expect OTOCs to grow.

One simple deformation of the model is to modify the adjoint field correlator to decay in time by hand. 
For example, one can modify eq.~\eqref{ktherm} into 
\begin{align}
K(\omega) = \frac{i}{1-y}\left(\frac{1}{\omega^2- (m -i \Gamma)^2} - \frac{y}{\omega^2-(m+i\Gamma)^2}\right)~.
\end{align}
where nonzero $\Gamma$ induces the adjoint correlator to decay as $\sim \exp(- \Gamma t)$.  
However, such a modification of the adjoint correlator changes the SD equation from eq.~\eqref{SDeq29} into  
\begin{align}
{G}(\omega) = {G}_0(\omega) - \frac{\lambda}{2 m (1- y)} {G}_0(\omega) {G}(\omega) \Bigl( G(\omega - m + i \Gamma ) + y \, G(\omega + m + i \Gamma) \Bigr)  \,,
\end{align}
and thus, we need to solve in complex frequency $\omega$ space, which is more complicated.

Even if we leave aside the issue of only partial thermalization in the IP model,
another crucial difference between the IP matrix model and the SYK model is the existence of a conformal fixed point in IR. To remove a dimensionful parameter, $m$, the adjoint mass in the IP model, we could replace the adjoint with a conformal matter field. 
It would be interesting to investigate the modification of the model, especially the free matrix sector while maintaining the solvability but to make the OTOCs grow. 

Finally, we point out that our results of the absence of OTOC growth imply that 
the relations between the out-of-time ordered correlators (four-point function) and the Krylov complexity are consistent with that the Krylov complexity gives the upper bounds for OTOCs.   In \cite{Iizuka:2023pov, Iizuka:2023fba}, we studied the Krylov complexity of the creation operator $a^\dagger_i$ for the fundamental field and found that it grows exponentially in time. Note that the Krylov complexity of the adjoint field $X_{ij}$ in the IP model would not grow, because the spectrum of the adjoint is discrete and they are free in the large $N$ limit. Our results of the OTOCs for the fundamental field imply that the Lyapunov coefficients for the fundamental field vanish in the parameter range we studied, and certainly, this is consistent with the nonzero Krylov complexity as an upper bound.


\acknowledgments
The work of NI was supported in part by JSPS KAKENHI Grant Number 18K03619 and also by MEXT KAKENHI Grant-in-Aid for Transformative Research Areas A ``Extreme Universe'' No.~21H05184. M.N.~was supported by the Basic Science Research Program through the National Research Foundation of Korea (NRF) funded by the Ministry of Education (RS-2023-00245035).

\appendix
\section{Numerical evaluation of eq.~\eqref{finalmasslessOTOCs} using the Green function $G(\omega)$ for $m > 0$}\label{app1}
In this appendix, we evaluate the OTOCs formula eq.~\eqref{finalmasslessOTOCs} with a finite number of rung cutoff, $n_{max}$, 
\begin{align}\label{NaiveOTOC}
\sum_{n=1}^{n_{max}}N {G}^{(n)}_4 (0, 0, t, -t) 
= \sum_{n=1}^{n_{max}} \left( \frac{\nu_T}{2 } \right)^{2n} 2^n \,  \left| \int  \frac{d\omega}{2\pi}  \left( G(\omega) \right)^{n+1} e^{ -i \omega t } \right|^2
\end{align}
using the the dressed Green function $G(\omega)$ obtained for $m \neq 0$. Note that this is {\it not} consistent physically. This is because the formula eq.~\eqref{finalmasslessOTOCs} and \eqref{NaiveOTOC} is obtained in the limit $m \to 0$. On the other hand, the dressed Green function $G(\omega)$ we use in \eqref{NaiveOTOC} is the one obtained for nonzero $m$. 

For the true OTOCs, we need to use the formula eq.~\eqref{finalOTOCs} instead of eq.~\eqref{finalmasslessOTOCs}, eq.~\eqref{finalmasslessOTOCs} is the $m \to 0$ limit of eq.~\eqref{finalOTOCs}. The reason for this evaluation is to show that, given the Green function  $G(\omega)$ for the case of nonzero mass $m > 0$, as we have shown in section \ref{mainnumerics}, even though \eqref{finalOTOCs} does not show exponential growth, eq.~\eqref{finalmasslessOTOCs} does show exponential growth by using the Green function $G(\omega)$ for $m > 0$. Thus, the explicit mass $m$ dependence between formulas \eqref{finalOTOCs} and \eqref{finalmasslessOTOCs} are so crucial that the behaviors significantly change. 

\begin{figure}[t]
\centering
     \begin{subfigure}[b]{0.6\textwidth}
         \centering
         \includegraphics[width=\textwidth]{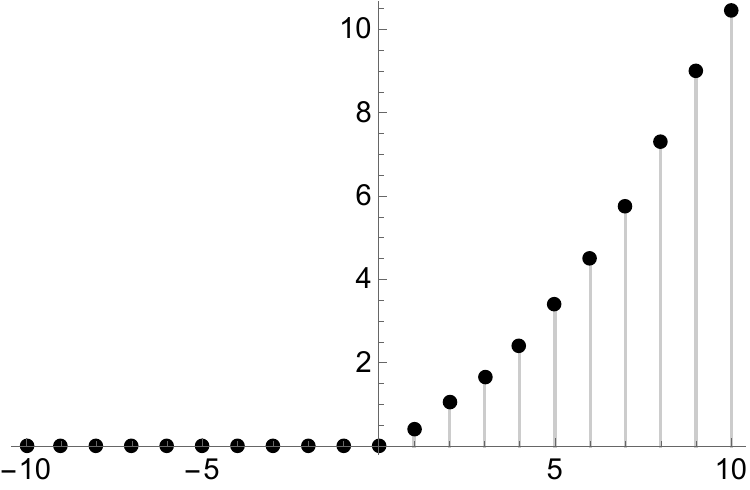}
       \put(5,15){$t$}
    \put(-200,190){$\text{Re}[\sum_{n=1}^{n_{max}}N {G}^{(n)}_4 (0, 0, t, -t)]$}
\label{fig:NaiveOTOC1}
\caption{$n_{max}=5$}
\vspace{10mm}
     \end{subfigure}
      \quad\quad\quad
     \begin{subfigure}[b]{0.6\textwidth}
         \centering
         \includegraphics[width=\textwidth]{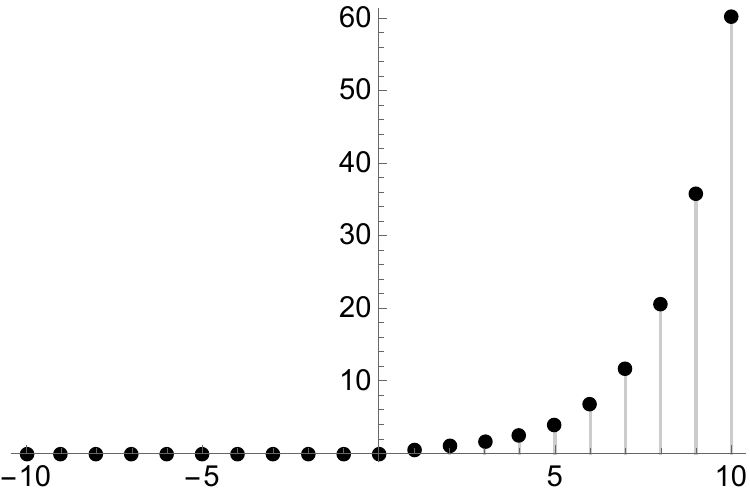}
          \put(5,15){$t$}
     \put(-200,190){$\text{Re}[\sum_{n=1}^{n_{max}}N {G}^{(n)}_4 (0, 0, t, -t)]$}
\caption{$n_{max}=10$}
\label{fig:NaiveOTOC2}
     \end{subfigure}
             \caption{Numerical plots of eq.~\eqref{NaiveOTOC} with the Green function $G(\omega)$ for $m=0.8$, $y=1$, $\nu_T=1$, $n_{max} = 5$ and $10$ case. These figures show manifestly exponential growth.}
        \label{fig:NaiveOTOC}
\end{figure}

To make the argument simple, let us consider the case of $y=1$, an infinite temperature limit. 
Figure \ref{fig:NaiveOTOC} shows numerical plots of $\text{Re}[\sum_{n=1}^{n_{max}}N {G}^{(n)}_4 (0, 0, t, -t)]$ with the Green function $G(\omega)$ obtained for $m=0.8$ case with $\nu_T=1$. One can see obvious growth of (\ref{NaiveOTOC}) at $t>0$.


For $m =0.8$, $y=1$, we evaluate the real OTOC using eq.~\eqref{finalOTOCsbeta2}  in Figure \ref{fig:OTOCm08v1y1}\footnote{Note that at the infinite temperature limit, $y=1$, eq.~\eqref{finalOTOCsbeta2} becomes identical to eq.~\eqref{finalOTOCs}.}, and we see that the real OTOC does not grow exponentially. On the other hand, here we see that eq.~\eqref{finalmasslessOTOCs} grows using the same Green function. Thus the manifest mass dependence in these formula are so crucial that in one case it gives exponential growth but in the other case, it does not grow.  


\section{$n_{max}$-dependence of the OTOCs with fixed time $t$}\label{app2}
In Figures \ref{fig:OTOCm08v1} and \ref{fig:OTOCm08v2}, the OTOCs $\text{Re}[N {G}^{n_{max}}_4 (0, -i\frac{\beta}{2}, t, - t -i\frac{\beta}{2})]$ for $m=0.8$, $\nu_T=1$ have a peak at $t \sim15.5$. For $y=0.4$, $0.7$, $1$, even though $n_{max}$ is set to $n_{max}=10$, the numerical values of $\text{Re}[N {G}^{n_{max}}_4 (0, -i\frac{\beta}{2}, t, - t -i\frac{\beta}{2})]$ at $t \sim15.5$ do not converge in our numerical computations. If we further increase $n_{max}$, the values of the peak at $t \sim15.5$ could be much larger.

\begin{figure}[t]
\centering
     \begin{subfigure}[b]{0.52\textwidth}
         \includegraphics[width=\textwidth]{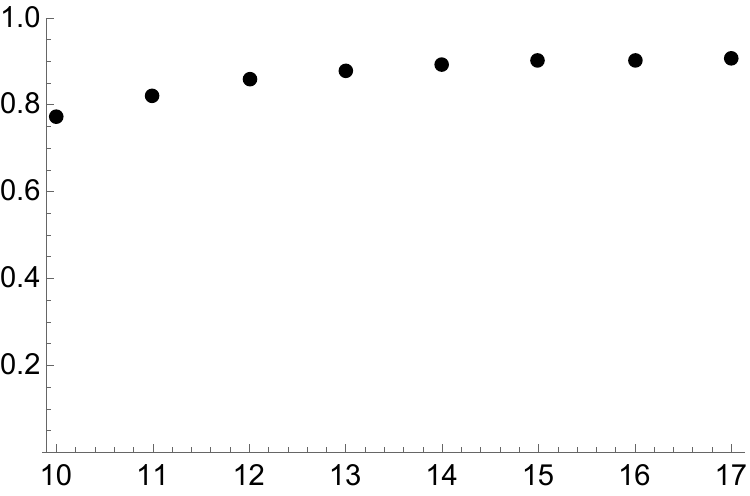}
       \put(5,15){$n_{max}$}
    \put(-260,165){$\text{Re}[N {G}^{n_{max}}_4 (0, -i\frac{\beta}{2}, t, - t -i\frac{\beta}{2})]$}
\label{fig:OTOCy04t15}
\caption{$y=0.4$}
\vspace{5mm}
     \end{subfigure}
     \begin{subfigure}[b]{0.52\textwidth}
         \centering
         \includegraphics[width=\textwidth]{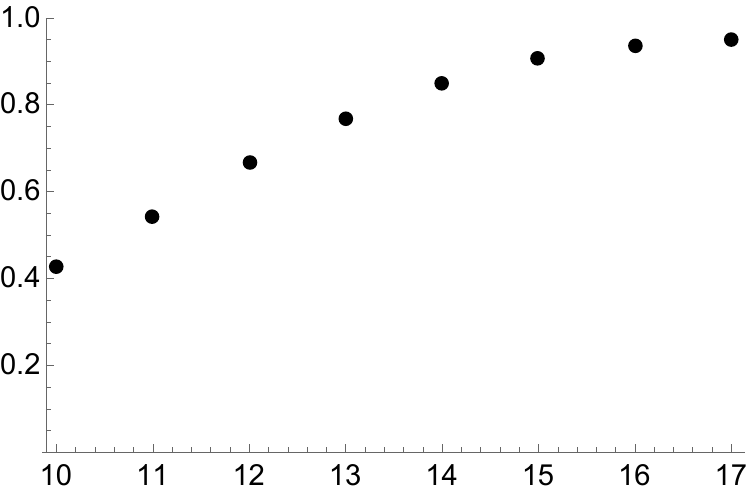}
          \put(5,15){$n_{max}$}
     \put(-260,165){$\text{Re}[N {G}^{n_{max}}_4 (0, -i\frac{\beta}{2}, t, - t -i\frac{\beta}{2})]$}
\caption{$y=0.7$}
\label{fig:OTOCy07t15}
\vspace{5mm}
     \end{subfigure}
          \begin{subfigure}[b]{0.52\textwidth}
         \centering
         \includegraphics[width=\textwidth]{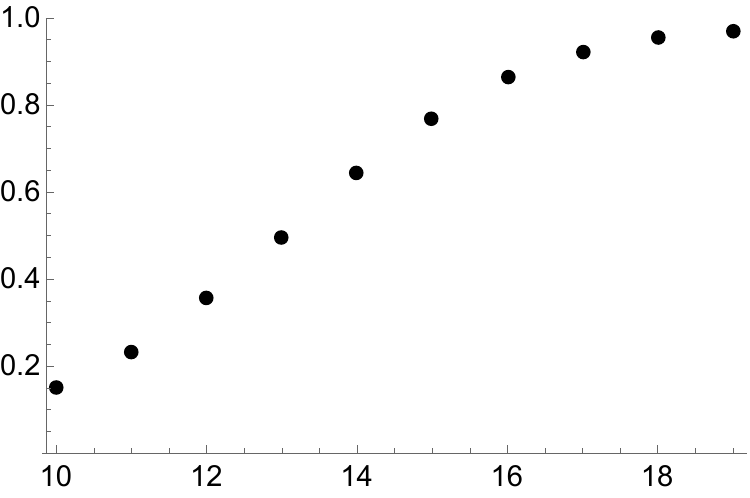}
       \put(5,15){$n_{max}$}
    \put(-260,165){$\text{Re}[N {G}^{n_{max}}_4 (0, -i\frac{\beta}{2}, t, - t -i\frac{\beta}{2})]$}
\label{fig:OTOCy1t15}
\caption{$y=1$}
     \end{subfigure}
             \caption{Numerical plots of the OTOCs $\text{Re}[N {G}^{n_{max}}_4 (0, -i\frac{\beta}{2}, t, - t -i\frac{\beta}{2})]$ at $t=15.5$ for $m=0.8$, $\nu_T=1$, $y=0.4$, $0.7$, $1$, where the horizontal axis is $n_{max}$.}
        \label{fig:OTOCt15}
\end{figure}

To verify it, we plot the $n_{max}$-dependence of $\text{Re}[N {G}^{n_{max}}_4 (0, -i\frac{\beta}{2}, t, - t -i\frac{\beta}{2})]$ at $t=15.5$ for $m=0.8, \nu_T=1$, $y=0.4$, $0.7$, $1$ in Figure \ref{fig:OTOCt15}. As $n_{max}$ increases, $\text{Re}[N {G}^{n_{max}}_4 (0, -i\frac{\beta}{2}, t, - t -i\frac{\beta}{2})]$ also increases as expected. However, the growth rate with respect to $n_{max}$ becomes small as $n_{max}$ increases, and the convergence of numerical values of $\text{Re}[N {G}^{n_{max}}_4 (0, -i\frac{\beta}{2}, t, - t -i\frac{\beta}{2})]$ at $t=15.5$ is better around $n_{max}\sim18$. These figures suggest that the numerical values approach $1$ as $n_{max}$ increases. 

\begin{figure}[t]
\centering
     \begin{subfigure}[b]{0.52\textwidth}
         \includegraphics[width=\textwidth]{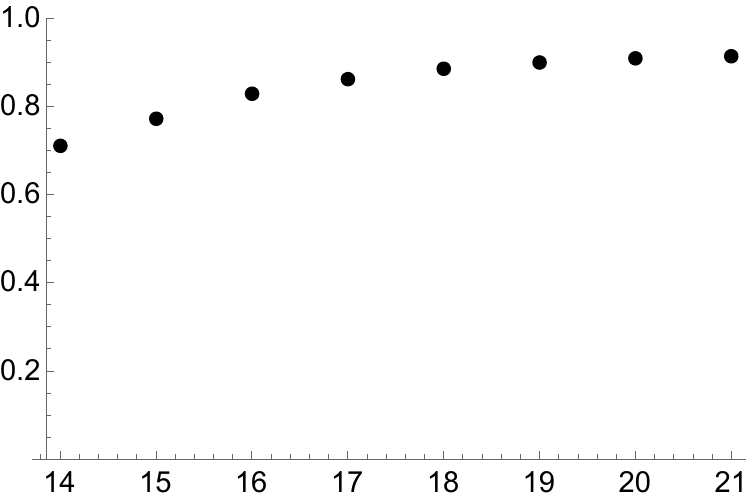}
       \put(5,15){$n_{max}$}
    \put(-260,165){$\text{Re}[N {G}^{n_{max}}_4 (0, -i\frac{\beta}{2}, t, - t -i\frac{\beta}{2})]$}
\label{fig:OTOCy04t235}
\caption{$y=0.4$}
\vspace{5mm}
     \end{subfigure}
     \begin{subfigure}[b]{0.52\textwidth}
         \centering
         \includegraphics[width=\textwidth]{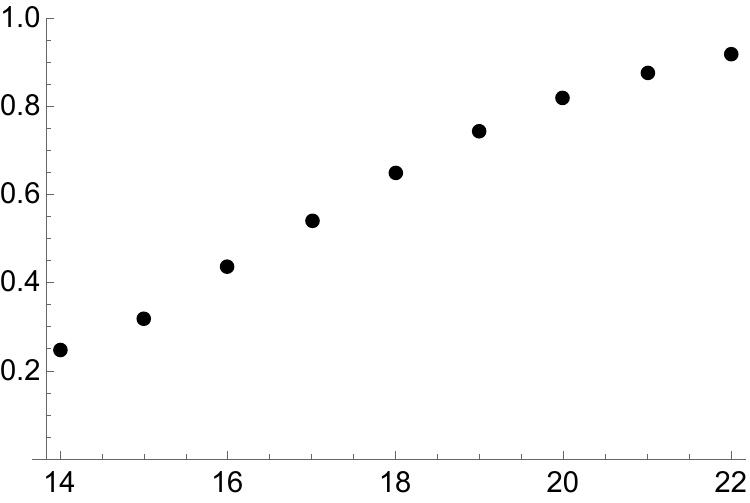}
          \put(5,15){$n_{max}$}
     \put(-260,165){$\text{Re}[N {G}^{n_{max}}_4 (0, -i\frac{\beta}{2}, t, - t -i\frac{\beta}{2})]$}
\caption{$y=0.7$}
\label{fig:OTOCy07t235}
\vspace{5mm}
     \end{subfigure}
          \begin{subfigure}[b]{0.52\textwidth}
         \centering
         \includegraphics[width=\textwidth]{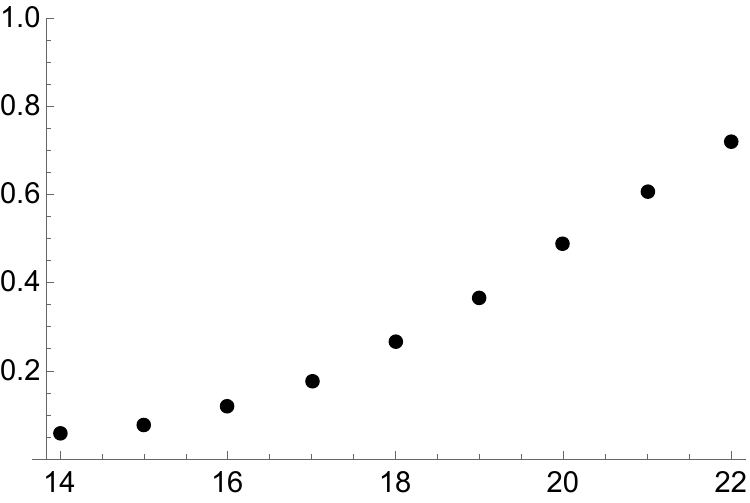}
       \put(5,15){$n_{max}$}
    \put(-260,165){$\text{Re}[N {G}^{n_{max}}_4 (0, -i\frac{\beta}{2}, t, - t -i\frac{\beta}{2})]$}
\label{fig:OTOCy1t235}
\caption{$y=1$}
     \end{subfigure}
             \caption{Numerical plots of the OTOCs $\text{Re}[N {G}^{n_{max}}_4 (0, -i\frac{\beta}{2}, t, - t -i\frac{\beta}{2})]$ at $t=23.5$ for $m=0.8$, $\nu_T=1$, $y=0.4$, $0.7$, $1$, where the horizontal axis is $n_{max}$.}
        \label{fig:OTOCt235}
\end{figure}

We also do a similar plot at $t=23.5$ for the next peak in Figure \ref{fig:OTOCt235}. For $y=0.4$, the convergence of $\text{Re}[N {G}^{n_{max}}_4 (0, -i\frac{\beta}{2}, t, - t -i\frac{\beta}{2})]$ at $t=23.5$ is better around $n_{max}\sim20$. However, for $y=0.7$ and $y=1$, the convergence is not so good for numerical computations up to $n_{max}=22$.

\newpage

\clearpage

\newpage

\clearpage

\bibliography{Ref}
\bibliographystyle{JHEP}

\end{document}